\documentclass[12pt,twoside]{amsart}
\usepackage{amssymb}
\usepackage[english]{babel}
\usepackage[utf8]{inputenc}
\usepackage{algorithm}
\usepackage[noend]{algpseudocode}
\usepackage{natbib}
\usepackage{comment} 
\usepackage{tikz}
\usetikzlibrary{calc}
\usepackage{kantlipsum} 

\usepackage{lineno}
\usepackage{lipsum}
\def\RR{\mathbb{R}}    
\def\PP{\mathbb{P}}    
\def\EE{\mathbb{E}}    
\def\P{{\mathcal P}}   
\def\R{{\mathcal R}}   
\def\G{{\mathcal G}}   
\def\R{\mathcal{R}}   

\def\proof{\noindent{\em Proof.}~}
\def\eproof{\mbox{\ }\hfill$\square$}
\newtheorem{theorem}{Theorem}
\newtheorem{lemma}{Lemma}
\newtheorem{proposition}{Proposition}
\newtheorem{corollary}{Corollary}
\newtheorem{assumption}{Assumption}
\newtheorem{definition}{Definition}
\newtheorem{example}[theorem]{\textsc{Example}}
\date{}

\title{Learning in Random Utility Models Via\\  Online Decision Problems }

\author{Emerson Melo}

\date{\today}

\calclayout
\calclayout
\thanks{Department of Economics,
	Indiana University, Bloomington, IN 47408, USA. Email: {\tt emelo@iu.edu}. I am very grateful to Bob Becker, Austin Knies, Jorge Lorca, and Pablo Pincheira  for their  valuable comments and suggestions that have greatly improved the paper. I am also very grateful to the Associate Editor and the anonymous referee for their insightful comments, which greatly improved the quality of the paper.}
\begin{document}

\maketitle

\pagestyle{myheadings} \thispagestyle{plain} \markboth{ }{ }
\begin{abstract} 
This paper examines the Random Utility Model (RUM) in repeated stochastic choice settings where decision-makers lack full information about payoffs. We propose a gradient-based learning algorithm that embeds RUM into an online decision-making framework. Our analysis establishes Hannan consistency for a broad class of RUMs, meaning the average regret relative to the best fixed action in hindsight vanishes over time. We also show that our algorithm is equivalent to the Follow-The-Regularized-Leader (FTRL) method, offering an economically grounded approach to online optimization. Applications include modeling recency bias and characterizing coarse correlated equilibria in normal-form games.

	\end{abstract}
%
\vspace{3ex}

\thispagestyle{empty}

\newcommand{\spacing}[1]{\renewcommand{\baselinestretch}{#1}\large\normalsize}
\textwidth      5.95in \textheight 600pt
\spacing{1.0}
 \newpage
\section{Introduction}\label{s1}

The Random Utility Model (RUM), introduced by \citet{Marschak1959}, \citet{BlockMarschak1959}, and \citet{BeckerGordonDegroot1963}, is the standard framework for modeling stochastic discrete choice in economics.\footnote{See \citet{Tversky1969} for early empirical evidence, and \citet{AgranovOrtoleva2016} for recent experimental support of stochastic choice in repeated decisions.} The seminal work of \citet{McFadden1978, mcffadens1, mcf1} established the economic and econometric foundations of the RUM, making it applicable to complex environments and welfare analysis (\citet{McFadden2001}, \citet{train_2009}).

In the classical RUM, a decision-maker (DM) chooses the option with the highest utility, where each utility is the sum of a deterministic component and a random shock. Choice probabilities reflect the likelihood of each option being utility-maximizing. Different assumptions about the shock distribution yield different choice models, with the goal of ensuring consistency with random utility maximization (\citet{mcf1}).

This framework relies on two strong assumptions: (i) a static decision environment and (ii) full knowledge of utilities. These rarely hold in dynamic settings, where DMs make repeated choices under uncertainty and face informational frictions such as limited attention or cognitive constraints.\footnote{See \citet{Matejka2015}, \citet{CaplinDean2015}, \citet{Caplinetal2019}, \citet{Fosgerauetal2019a}, and \citet{Natenzon2019}. \citet{WEBB2019} develops a dynamic RUM based on bounded evidence accumulation. \citet{cerreiavioglio2021multinomial} axiomatize the MNL under time-constrained information processing.}

We propose a novel extension of the Random Utility Model (RUM), termed the RUM-ODP model, which integrates discrete choice into an Online Decision Problem (ODP) framework. In this setting, the decision maker (DM) selects a probability distribution over actions in each period without knowing the true deterministic utilities, which must be inferred from noisy, feedback-driven learning. Utilities are updated cumulatively and imperfectly, so the DM’s probabilistic choices reflect evolving beliefs shaped by random shocks. Thus, at each time step, the DM’s behavior can be viewed as a RUM with utilities estimated imperfectly from past experience.

In this setting, the DM minimizes regret—the gap between actual cumulative utility and the best fixed action in hindsight (\citet{Bell1982}, \citet{LoomesSugden1982}, \citet{FISHBURN198231}). A learning process with vanishing average regret is said to be Hannan consistent (\citet{Hahn1957}). This property, often referred to as no-regret—a, plays  a central role  in both economics and machine learning (\citet{BianchiLugosi2003}, \citet{Roughgarden2016}).

The RUM-ODP applies to a wide range of problems, including repeated consumer choice, prediction with expert advice, and online learning in adversarial or strategic settings (\citet{Hazan2017}). It can also be interpreted as a repeated game between the DM and nature, where regret measures performance.\footnote{See \citet{LittlestoneWarmuth1994}, \citet{FreundSchapire1997, FreundSchapire1999}, \citet{FosterVohra1999}, and \citet{CesaBianchiLugosi2006}.}

A key advantage of the RUM-ODP is its minimal informational assumptions. Unlike Bayesian models (e.g., \citet{Matejka2015}, \citet{Fosgerauetal2019a}), it requires no prior distribution or learning rule—only observed feedback from past actions.

We make four main contributions. First, we introduce the Social Surplus Algorithm (SSA) as a behavioral foundation for learning under the RUM-ODP model. SSA is based on the fact that RUM choice probabilities equal the gradient of the Social Surplus function.\footnote{Defined as $\varphi(\bold{v}) = \mathbb{E}_\epsilon[\max_{i \in A}\{\bold{v}_i + \epsilon_i
\}]$, where $\bold{v}_i$ is deterministic utility and $\epsilon_i$ is a preference shock; choice probabilities are given by $\partial \varphi(\bold{v})/\partial \bold{v}_i$.} We show that SSA is Hannan consistent across a broad class of RUMs, including those with correlated, nested, or grouped alternatives.

Second, we apply SSA to the Generalized Extreme Value (GEV) family—including the multinomial logit (MNL), nested logit (NL), paired combinatorial logit (PCL), generalized nested logit (GNL), and  ordered GEV (OGEV)—and show it ensures Hannan consistency. This opens up new possibilities for studying learning and substitution patterns in dynamic demand environments previously unexplored for GEV models.

Third, we establish a duality between SSA and the Follow-The-Regularized-Leader (FTRL) algorithm, removing the need to specify a regularization term explicitly. This connection gives a behavioral interpretation of FTRL via RUMs and provides a recursive structure for FTRL-based choice rules across many RUMs, generalizing the Exponential Weights Algorithm (EWA) beyond the MNL case. We also extend the model to include recency bias and show it remains Hannan consistent, grounding the Optimistic FTRL algorithm.

Finally, we apply our framework to no-regret learning in normal-form games. Compared to potential-based dynamics, our RUM-ODP model offers a flexible and behaviorally grounded alternative for analyzing strategic behavior.

The paper is organized as follows: Section \ref{s2} introduces the model and SSA. Section \ref{s33} studies the GEV case. Section \ref{s4} analyzes the FTRL algorithm and its relationship to SSA. Section \ref{S5_Games} explores no-regret learning in games. Section \ref{s5} reviews related work, and Section \ref{s6} concludes. Proofs appear in Appendix \ref{Proofs}.

\noindent\textbf{Notation.} Let $\langle\cdot,\cdot\rangle$ denote the inner product between two vectors. For a convex function $f:K\subseteq\RR^n\longrightarrow\RR$, $\partial f(\bold{x})$ denotes the subgradient of $f$ at $\bold{x}.$  Let $\nabla f(\bold{x})$ denote the gradient of a function $f:K\subseteq\RR^N\longrightarrow\RR$ evaluated at point $\bold{x}$.  The $i$th element of $\nabla f(\bold{x})$ is denoted by $\nabla_if(\bold{x})$.  The Bregman divergence  associated to a function $f$ is given  by $D_f(\bold{x}||\bold{y})=f(\bold{y})-f(\bold{x})-\langle\nabla f(\bold{x}),\bold{y}-\bold{x}\rangle$.  The Hessian of  $f$ at point $\bold{x}$ is denoted by $\nabla^2f(\bold{x})$ with entries given by $\nabla^2_{ij}f(\bold{x})$ for $i,j=1,\ldots,N$. Let  $\|\cdot\|$ denote a norm in $\RR^N$ where $\|\cdot\|_*$ is its dual norm.  Let $\bold{A}\in\RR^{N\times N}$ denote a $N$-square matrix. We define the norm $\|\cdot\|_{\infty,1}$ associated to the matrix  $\bold{A}$ as $\|\bold{A}\|_{\infty,1}=\max_{\|\bold{v}\|_1\leq 1} \|\bold{A}\bold{v}\|_1$. Finally, the trace of a matrix $\bold{A}\in\RR^N$ is denoted by $Tr(\bold{A})$.  
\section{Online decision problems and  the Social Surplus Algorithm}\label{s2}

Let $A=\{1,\ldots,N\}$ be a finite set of alternatives. Let $\Delta_N$ denote the $N$-dimensional simplex over the set  $A$. Let $T\geq 2$ denote the (exogenous) number of periods. Let $\bold{u}_t=(\bold{u}_{1t},\ldots\bold{u}_{Nt})$ be a random vector, where $\bold{u}_{it}$ denotes the stochastic payoff associated to option $i\in A$ for $t=1,\ldots, T$. The realizations of the vector $\bold{u}_t$ are determined by the environment (nature), which in principle can be adversarial. We assume that the  vector $\bold{u}_t$ takes values on a compact set  $\mathcal{U}\subseteq \RR^N$. In particular,  we assume that   $\|\bold{u}_t\|_{\infty}\leq u_{\max}\quad\mbox{for all $t$.}$
\smallskip

 In this paper, we study the following ODP. At each time period $t = 1, \ldots, T$, the DM selects an action $\mathbf{x}_t \in \Delta_N$ based on the information available up to period $t-1$. After committing to $\mathbf{x}_t$, the DM observes the realization of the payoff vector $\mathbf{u}_t$ and receives an expected payoff of $\langle \mathbf{u}_t, \mathbf{x}_t \rangle$. The DM’s objective is to select a sequence of actions $\mathbf{x}_1, \ldots, \mathbf{x}_T$ that minimizes her regret, defined as the difference between the cumulative payoff of the best fixed decision in hindsight and the cumulative payoff actually received. 

\begin{definition}\label{Regret_Definition} 
Consider a time horizon of $T$ periods and a sequence of choices $\mathcal{A} = \{\mathbf{x}_1, \ldots, \mathbf{x}_T\}$, where each $\mathbf{x}_t \in \Delta_N$. The regret associated with the sequence $\mathcal{A}$ is defined as:
\begin{equation}\label{Regret_eq}
    \textsc{R}_{\mathcal{A}}^T = \max_{\mathbf{x} \in \Delta_N} \left\langle \boldsymbol{\theta}_T, \mathbf{x} \right\rangle - \sum_{t=1}^{T} \left\langle \mathbf{u}_t, \mathbf{x}_t \right\rangle,
\end{equation}
where $\boldsymbol{\theta}_T \triangleq \sum_{t=1}^{T} \mathbf{u}_t$ is the cumulative payoff vector up to period $T$.
\end{definition}

The regret $\textsc{R}_{\mathcal{A}}^T$ measures the performance gap between the DM’s sequence of actions $\mathbf{x}_1, \ldots, \mathbf{x}_T$ and the best fixed action in hindsight—that is, the optimal action $\mathbf{x}^* \in \arg\max_{\mathbf{x} \in \Delta_N} \langle \boldsymbol{\theta}_T, \mathbf{x} \rangle$ chosen with full knowledge of the cumulative payoff vector $\boldsymbol{\theta}_T$. This ideal benchmark assumes that the DM has access to the entire sequence of payoff vectors $\mathbf{u}_1, \ldots, \mathbf{u}_T$ before making any decisions.

In contrast, within the ODP framework, the DM selects actions sequentially, relying only on information observed up to the previous period. As such, the sequence $\mathcal{A}$ captures the DM’s learning dynamics over time and can be naturally associated with a particular learning algorithm.\footnote{Definition~\ref{Regret_Definition} can equivalently be written as
\[
\textsc{R}_{\mathcal{A}}^T = \max_{j \in A} \{ \boldsymbol{\theta}_{jT} \} - \sum_{t=1}^{T} \langle \mathbf{u}_t, \mathbf{x}_t \rangle,
\]
where $\boldsymbol{\theta}_{jT} \triangleq \sum_{t=1}^T \mathbf{u}_{jt}$ denotes the cumulative payoff of alternative $j$ up to time $T$.}
\smallskip

To formalize the idea that the DM employs learning algorithms aimed at minimizing regret over time, we introduce the concept of Hannan Consistency \citep{Hahn1957}.

\begin{definition}[\citet{Hahn1957}]\label{Hannan_Consistency}
A sequence of decisions $\mathcal{A}$ is said to be \emph{Hannan consistent} if the regret defined in Eq.~\eqref{Regret_eq} grows sublinearly with $T$, that is,
\begin{equation}\label{Hannan_consistency}
\textsc{R}_{\mathcal{A}}^T = o(T).
\end{equation}
\end{definition}
Intuitively, Definition~\ref{Hannan_Consistency} states that the sequence $\mathcal{A}$ is Hannan consistent if the \emph{average regret} associated with $\mathcal{A}$ vanishes as $T \to \infty$. Formally, this is equivalent to requiring that $\textsc{R}_{\mathcal{A}}^T$ grows sublinearly in $T$, i.e., Eq.\eqref{Hannan_consistency} holds. Thus, Hannan consistency ensures that a sequence of choices performs asymptotically as well as the best fixed strategy in hindsight, denoted by $\bold{x}^*$. Alternatively, when a sequence $\mathcal{A}$ satisfies condition\eqref{Hannan_consistency}, we say that the algorithm $\mathcal{A}$ has the \emph{no-regret} property.\footnote{Throughout this paper, we use the terms \emph{Hannan consistency} and \emph{no-regret learning} interchangeably.}
\smallskip

By combining the ODP model with the concept of regret, we can formulate learning algorithms that adhere to Definition \ref{Hannan_Consistency}. A common approach in game theory and online convex optimization (OCO) problems is to investigate the FTRL algorithm with the entropic penalty, as outlined in \citet{Hazan2017} and related literature. 
\smallskip

This paper shows that a broad class of discrete choice models satisfy Hannan consistency, extending no-regret learning analysis beyond the traditional MNL model. Consequently, no-regret learning can be explored within a richer and more flexible class of behavioral models.

\subsection{The  Social Surplus Algorithm} In this section, we develop a simple algorithm that combines the ODP framework with the theory of RUMs (\citet[Ch. 5]{mcf1}). This integration extends the classic approach of \citet{Hahn1957}. Specifically, Hannan considers an ODP model in which the DM’s choice is given by:
\begin{equation}\label{FTPL1}
\tilde{\bold{x}}_{t+1} \in \arg\max_{\bold{x} \in \Delta_N} \langle \boldsymbol{\theta}_t + \eta \epsilon_{t+1}, \bold{x} \rangle \quad \text{for } t = 1, \ldots, T,
\end{equation}
where $\boldsymbol{\theta}_t$ denotes the cumulative payoff vector up to period $t$, $\epsilon_{t+1} = (\epsilon_{1t+1}, \ldots, \epsilon_{Nt+1})$ is a vector of random preference shocks, and $\eta > 0$ is a strictly positive parameter.

\smallskip

Equation (\ref{FTPL1}) defines a recursive choice rule in which the DM selects $\tilde{\bold{x}}_{t+1}$ by maximizing a perturbed payoff. The choice depends on the cumulative information $\boldsymbol{\theta}_t$, the realization of the shock vector $\epsilon_{t+1}$, and the learning parameter $\eta$. The parameter $\eta$ can also be interpreted as a measure of \emph{accuracy} in the DM’s choices: smaller values imply more deterministic behavior, while larger values reflect greater randomness.\footnote{This interpretation of $\eta$ as an accuracy parameter is also discussed in \citet{cerreiavioglio2021multinomial} within a different context.}

Intuitively, equation (\ref{FTPL1}) reflects a DM who samples a random realization of $\epsilon_{t+1}$ to smooth out her optimization problem. In the OCO literature, this method is known as Follow the Perturbed Leader (FTPL). A common assumption in this framework is that $\epsilon_{t+1}$ is drawn from an $N$-dimensional uniform distribution. The main advantage of the FTPL approach is that it can induce stability in the DM’s choices.\footnote{In the game theory literature, a similar method is referred to as fictitious stochastic play. For details, see \citet{FudenbergLevina1998}.} Indeed, \citet{Hahn1957} shows that when $\epsilon_t$ is i.i.d. following a  uniformly distribution, the sequence generated by solving problem (\ref{FTPL1}) satisfies Definition \ref{Hannan_Consistency}.

However, the regret analysis of FTPL relies heavily on probabilistic arguments tied to the stochastic structure of $\epsilon_{t+1}$, rather than on a general, unified framework (see \citet{Abernethyetal2016}). More importantly, from an economic perspective, the interpretation of the FTPL approach remains unclear, making it difficult to ground the repeated stochastic choice problem in a solid behavioral foundation.
\smallskip

In this section, we propose shifting the focus from particular realizations of $\epsilon_{t+1}$ to leveraging its entire distribution. Specifically, under reasonably general assumptions about the distribution of $\epsilon_{t+1}$, we can utilize the theory of RUMs to develop an alternative formulation of Hannan’s original FTPL algorithm. This approach offers a more structured foundation by relying on the statistical properties of the perturbations rather than individual draws. Throughout the paper, we adopt the following assumption:

\begin{assumption}\label{Shocks_Assumption}
For all $t \geq 1$, the random vector $\epsilon_t = (\epsilon_{1t}, \ldots, \epsilon_{Nt})$ follows a joint distribution $F = (F_1, \ldots, F_N)$ with zero mean, which is absolutely continuous with respect to the Lebesgue measure, independent of both $t$ and $\boldsymbol{\theta}_t$, and fully supported on $\RR^N$.
\end{assumption}

Assumption \ref{Shocks_Assumption} is standard in the literature on random utility and discrete choice models (see \citet[Ch. 5]{mcf1}). The conditions of full support and absolute continuity imply that problem (\ref{FTPL1}) can be equivalently rewritten as:
\begin{equation}\label{FTPL2}
\max_{\bold{x} \in \Delta_N} \langle \boldsymbol{\theta}_{t} + \eta \epsilon_{t+1}, \bold{x} \rangle = \max_{j \in A} \{ \boldsymbol{\theta}_{jt} + \eta \epsilon_{jt+1} \}, \quad \text{for } t = 1, \ldots, T.
\end{equation}

From expression (\ref{FTPL2}), it follows that Assumption \ref{Shocks_Assumption} ensures $\tilde{\bold{x}}_{t+1}$ is a corner solution—that is, the DM selects a single alternative with probability one. Moreover, since the maximum function is convex, and by defining $\tilde{\varphi}(\boldsymbol{\theta}_t + \eta \epsilon_{t+1}) \triangleq \max_{j \in A} \{ \boldsymbol{\theta}_{jt} + \eta \epsilon_{jt+1} \}$, the choice $\tilde{\bold{x}}_{t+1}$ can be characterized as:

\begin{equation}\label{FTPL3}
\tilde{\bold{x}}_{t+1} \in \partial \tilde{\varphi}(\boldsymbol{\theta}_{t} + \eta \epsilon_{t+1}),
\end{equation}

where $\partial \tilde{\varphi}(\cdot)$ denotes the subdifferential of the convex function $\tilde{\varphi}$. Thus, the optimal solution $\tilde{\bold{x}}_{t+1}$ corresponds to a subgradient of $\tilde{\varphi}(\boldsymbol{\theta}_t + \eta \epsilon_{t+1})$ (\citet{Rockafellar1970}).
\smallskip

  A natural way to extend (\ref{FTPL3}) is to incorporate the full distribution of $\epsilon_{t+1}$ by taking the expectation of $\tilde{\varphi}(\boldsymbol{\theta}_t + \epsilon_{t+1})$. Formally, define the function $\varphi: \RR^N \to \RR$ as:
\begin{equation}\label{SS_Function}
\varphi(\boldsymbol{\theta}_t) \triangleq \EE \left[ \max_{j \in A} \{ \boldsymbol{\theta}_{jt} + \eta \epsilon_{jt+1} \} \right],
\end{equation}
where the expectation is taken with respect to the distribution $F$.

In the RUM literature, $\varphi(\boldsymbol{\theta}_t)$ is referred to as the social surplus function, as it captures the aggregate effect of the distribution $F$. Importantly, $\varphi(\boldsymbol{\theta}_t)$ is both convex and differentiable on $\RR^N$. This differentiability implies that the choice probability vector $\bold{x}_{t+1}$ can be characterized as the gradient of $\varphi(\boldsymbol{\theta}_t)$:
\begin{equation}\label{SS_Function_gradient}
\nabla \varphi(\boldsymbol{\theta}_t) = \bold{x}_{t+1}, \quad \text{for } t = 1, \ldots, T-1.
\end{equation}

This result follows from the well-known Williams-Daly-Zachary (WDZ) theorem (see \citet[p. 3104]{Rust1994}), which establishes a connection between choice probabilities and the gradient of the expected maximum utility in the context of RUMs.
 
 \smallskip
 
 From an economic perspective, the social surplus function provides a RUM-based interpretation of the DM’s choices. The cumulative payoff vector $\boldsymbol{\theta}_t$ serves as a proxy (or estimate) of the unknown utility vector $\bold{u}_t$, with $\boldsymbol{\theta}_{jt}$ reflecting the historical performance of alternative $j$. The shock $\epsilon_{jt+1}$ represents a random preference component affecting the DM’s perception of this alternative. Applying this logic across all $j \in A$, the DM’s stochastic choices align with the theory of RUMs.

Crucially, the distributional assumptions on $\epsilon_{t+1}$ determine the form of the resulting choice rule, enabling the analysis of models that capture similarity and correlation among alternatives—such as probit or nested logit. We refer to this framework, which links RUMs and  ODPs, as the RUM-ODP model.

The following proposition is a direct consequence of  expression (\ref{SS_Function_gradient}).
\begin{proposition}\label{SS_Equiv}
Let Assumption \ref{Shocks_Assumption} hold. Then, for all $t = 1, \ldots, T$,
$$\EE[\tilde{\bold{x}}_{t+1}]= \bold{x}_{t+1}.$$
\end{proposition}

 This result implies the gradient of the social surplus function can be expressed as:
\begin{eqnarray}\label{FTPL_Potential}
\nabla\varphi(\boldsymbol{\theta}_t)&=&\EE\left[\arg\max_{\bold{x}\in\Delta_N}\langle\boldsymbol{\theta}_t+\eta\epsilon_{t+1},\bold{x}\rangle\right],\\
&=&\left(\PP\left(i=\arg\max_{j\in A}\{\boldsymbol{\theta}_{jt}/\eta+\epsilon_{jt+1}\}\right)\right)_{j\in A},\nonumber\\
&=& \left(\PP(\boldsymbol{\theta}_{jt}/\eta+\epsilon_{jt}\geq\boldsymbol{\theta}_{it}/\eta+\epsilon_{it},\forall j\neq i)\right)_{j\in A}.\nonumber
\end{eqnarray}

Additionally, the social surplus function satisfies the scaling properties:
$\varphi(\boldsymbol{\theta})=\eta\varphi(\boldsymbol{\theta}/\eta)$ and $\nabla\varphi(\boldsymbol{\theta})=\nabla\varphi(\boldsymbol{\theta}/\eta)$, which will be used in subsequent analysis.
\smallskip

Importantly, the convexity of $\varphi$ allows us to construct a learning algorithm, which we call the Social Surplus Algorithm (SSA). To proceed, we impose the following technical condition on the Hessian of $\varphi(\boldsymbol{\theta}_t)$.

\begin{assumption}\label{Gradient_LL}
For all $t \geq 1$, the Hessian of the social surplus function satisfies:
$$2\,\text{Tr}(\nabla^2 \varphi(\boldsymbol{\theta}_t)) \leq \frac{L}{\eta},$$
for some constant $L > 0$.
\end{assumption}

The previous assumption is nonstandard, as it places a condition on the trace of the Hessian of $\varphi(\boldsymbol{\theta}_t)$. This condition guarantees that the gradient of $\varphi$ is Lipschitz continuous, as shown below:

\begin{lemma}\label{Gradient_Lipschitz}
Let Assumptions \ref{Shocks_Assumption} and \ref{Gradient_LL} hold. Then the gradient $\nabla \varphi(\boldsymbol{\theta})$ is Lipschitz continuous with constant ${L}/{\eta}$:
\[
\|\nabla \varphi(\boldsymbol{\theta}) - \nabla \varphi(\tilde{\boldsymbol{\theta}})\|_1 \leq \frac{L}{\eta} \|\boldsymbol{\theta} - \tilde{\boldsymbol{\theta}}\|_1, \quad \forall\, \boldsymbol{\theta}, \tilde{\boldsymbol{\theta}} \in \RR^N.
\]
\end{lemma}

Recall that $\varphi(\boldsymbol{\theta}) = \eta \varphi(\boldsymbol{\theta}/\eta)$. Using this identity, Lemma \ref{Gradient_Lipschitz} also implies that the scaled function $\eta \varphi(\boldsymbol{\theta}/\eta)$ has an $L$-Lipschitz continuous gradient:
$$\|\nabla \varphi(\boldsymbol{\theta}/\eta) - \nabla \varphi(\tilde{\boldsymbol{\theta}}/\eta)\|_1 \leq L \|\boldsymbol{\theta}/\eta - \tilde{\boldsymbol{\theta}}/\eta\|_1.$$

Although technical, Assumption \ref{Gradient_LL} holds for many well-known RUMs, including the MNL, nested logit, and GEV models, as shown in Section~\ref{s33}.

\smallskip

We are now ready to introduce the SSA.
\newpage

\begin{algorithm}
	\caption{Social Surplus Algorithm}\label{alg:euclid1}
	\begin{algorithmic}[1]
		\State Input: $\eta>0$, $F$ a distribution on $\RR^N$, and $\Delta_N$.
		\State Let $\boldsymbol{\theta}_0=0$ and choose  $\bold{x}_1=\nabla\varphi(\bold{0})$ 
		\State $\bold{for}$ $t=1$ to $T$ \textbf{do}
		\State The DM chooses $\bold{x}_{t}=\nabla\varphi(\boldsymbol{\theta}_{t-1})$
		\State The environment reveals  $\bold{u}_{t}$
		\State The DM receives the payoff $\langle\bold{u}_{t}, \nabla\varphi(\boldsymbol{\theta}_{t-1})\rangle$
		\State Update $\boldsymbol{\theta}_{t}=\bold{u}_t+\boldsymbol{
		\theta}_{t-1}$ and choose
		$$\bold{x}_{t+1}=\nabla\varphi(\boldsymbol{\theta}_t)$$
		\State \textbf{end for}
	\end{algorithmic}
\end{algorithm} 

The proposed algorithm builds on $\varphi(\boldsymbol{\theta}_t)$ and its gradient, resembling a specific instance of the Gradient-Based Algorithm in \citet{Abernethyetal2016}. However, our focus diverges: we aim to uncover the economic foundations of Algorithm 1 by linking it to a broad class of RUMs, including NL, GEV, and OGEV. Unlike prior work, our analysis emphasizes correlation structures and nesting patterns among alternatives, and we derive closed-form expressions for $\nabla\varphi$, offering insights beyond those in \citet{Abernethyetal2016}.

The use of gradient-descent methods in repeated choice problems traces back to \citet{Blackwell_1956}. This line of work was extended by \citet{BianchiLugosi2003} for sequential decisions, and by \citet{HART_JET_200126, HART_GEB_2003375} through $\Lambda$-strategies, using potential functions.

A natural question is whether results from \citet{BianchiLugosi2003} apply to the RUM-ODP model. Doing so requires $\varphi$ to be additive, which holds only if the shocks $\epsilon_{it}$ are i.i.d.—as in the MNL model. Most RUMs we consider, however, do not meet this condition. Similarly, \citet{HART_JET_200126, HART_GEB_2003375} require the gradient $\nabla\varphi$ to vanish over the approachable set, which assumes bounded support for $\epsilon_t$ and excludes standard RUMs. Hence, their results don’t directly apply to our setting.

A further distinction lies in the interpretation of the potential function. In the aforementioned works, it measures regret. In contrast, in our framework, the social surplus function $\varphi$ not only tracks regret but also represents the expected utility under choice probabilities $\bold{x}_t$. Specifically, defining
\[ e_{jt}(\boldsymbol{\theta}_{t-1})\triangleq \EE[\epsilon_{jt}\mid j=\arg\max_{k\in A}{\boldsymbol{\theta}_{kt-1}+\epsilon_{kt}}], \]
we express $\varphi(\boldsymbol{\theta}_{t-1})$ as:
\begin{equation}\label{Potential_Econ_concise}
\varphi(\boldsymbol{\theta}_{t-1}) = \sum_{j=1}^{N} \bold{x}_{jt} \left(\boldsymbol{\theta}_{jt-1} + \eta e_{jt}(\boldsymbol{\theta}_{t-1})\right),
\end{equation}
which highlights the role of $\epsilon_t$ in shaping $\varphi$. Different distributions of $\epsilon_t$ yield different surplus functions, allowing the SSA to model a wide range of discrete choice scenarios with varying degrees of correlation. Notably, the SSA accommodates settings like NL  and GNL models and can be implemented using closed-form gradients within the GEV family, including MNL as a special case.

We now turn to the main result of this section.

\begin{theorem}\label{SS_Surplus_algorithm_regret}
Let Assumptions \ref{Shocks_Assumption} and \ref{Gradient_LL} hold. Then, the regret of the SSA satisfies:
\begin{equation}\label{Bound1}
\textsc{R}_{SSA}^T \leq \eta \varphi(\bold{0}) + \frac{L}{2\eta} T u_{\max}^2.
\end{equation}
Moreover, choosing the optimal step size $\eta = \sqrt{\frac{L T u_{\max}^2}{2\varphi(\bold{0})}}$ yields the bound:
\begin{equation}\label{Bound2}
\textsc{R}_{SSA}^T \leq u_{\max} \sqrt{2\varphi(\bold{0}) L T}.
\end{equation}
\end{theorem}

Theorem \ref{SS_Surplus_algorithm_regret} shows that the SSA achieves Hannan consistency for a broad class of RUMs. This result highlights the role of the parameters $L$, $\eta$, and the social surplus function evaluated at $\bold{u} = \bold{0}$, where $\eta \varphi(\bold{0}) = \eta \mathbb{E}(\max_{i=1,\ldots,N} \epsilon_i)$. As a consequence, the SSA can be implemented using a wide range of stochastic choice models, depending on the specific functional form of $\nabla \varphi(\boldsymbol{\theta}_t)$. In Section \ref{s33}, we will illustrate how this general result applies to different RUM specifications.

The proof of Theorem \ref{SS_Surplus_algorithm_regret} builds on the convexity of the  function $\varphi(\boldsymbol{\theta}_t)$. In particular, it relies on convex duality arguments to derive regret bounds. 


In the special case of the widely used MNL model—where $\epsilon_t$ follows an extreme value type I distribution—the bound in Theorem \ref{SS_Surplus_algorithm_regret} simplifies to:
\begin{corollary}\label{Cor_Logit}
Assume that for all $t$, the random shock vector $\epsilon_t$ follows an Extreme Value Type I distribution:
\begin{equation}\label{EV1}
F(\epsilon_{1t},\ldots,\epsilon_{Nt}) = \exp\left(-\sum_{j=1}^N e^{-\epsilon_{jt}}\right) \quad \forall t.
\end{equation}
Then, under the SSA, choosing $\eta = \sqrt{Tu^2_{\max} / (2 \log N)}$ yields the regret bound:
\[
\textsc{R}_{SSA}^T \leq u_{\max} \sqrt{2 \log N \cdot T}.
\]
\end{corollary}

This result follows from the well-known expression for the social surplus function in the MNL model:
$$\varphi(\boldsymbol{\theta}_t) = \eta \log\left( \sum_{j=1}^N e^{\boldsymbol{\theta}_{jt}/\eta} \right),$$
implying $\eta \varphi(\bold{0}) = \eta \log N$. The gradient $\nabla \varphi(\boldsymbol{\theta})$ corresponds to the softmax function and is $1/\eta$-Lipschitz continuous. Substituting these into Theorem \ref{SS_Surplus_algorithm_regret} yields the stated bound.

Behaviorally, the MNL model assumes that the shocks $\{\epsilon_{jt+1}\}_{j \in A}$ are independent across alternatives. While analytically convenient, this assumption may be too restrictive in settings where option payoffs exhibit correlation. The next section explores how the SSA can be extended to accommodate more flexible RUMs that relax the independence assumption.
 \section{The SSA and GEV models}\label{s33}
 This section establishes a connection between the RUM-ODP model, the SSA approach, and the class of GEV models. To our knowledge, the GEV framework has not yet been explored in the context of no-regret learning.

\smallskip

Originally introduced by \citet{McFadden1978}, the GEV class extends the MNL model by accommodating flexible patterns of dependence among the unobserved components of utility. Despite this added complexity, it retains analytical tractability through closed-form expressions for choice probabilities.

\smallskip

A key concept in McFadden’s development of the GEV class is the \emph{generator function}, defined as follows.
 \begin{definition}\label{Generator_GEV} A  function $G:\RR^N_{+}\longrightarrow\RR_+$ is a generator if the following conditions hold:
 	\begin{itemize}
 		\item[i)]Non-negativity: For all $\bold{y}=(\bold{y}_1,\ldots,\bold{y}_N)\in \RR^N_+$, $G(\bold{y})\geq 0$.
 		\item[ii)]Homogeneity of degree 1:  $G(\lambda\mathbf{y})=\lambda G(\bold{y})$ for all $\bold{y}\in \RR^N_+$ and $\lambda>0.$
 		\item[iii)]Coercivity in each argument: For each $j = 1, \ldots, N$, it holds that $G(\bold{y}) \to \infty$ as $\mathbf{y}_j \to \infty$, with all other components of $\mathbf{y}$ held fixed.
 		\item[iv)]   Sign structure of cross-partial derivatives: For any set of $k$ distinct indices $j_1, \ldots, j_k \in \{1, \ldots, N\}$, the $k$-th order mixed partial derivative satisfies:
$$\frac{\partial^k G(\bold{y})}{\partial \mathbf{y}_{j_1} \cdots \partial \mathbf{y}_{j_k}} \begin{cases}
\geq 0 & \text{if } k \text{ is odd}, \\
\leq 0 & \text{if } k \text{ is even}.
\end{cases}$$
 	\end{itemize} 
 \end{definition}

 \citet{McFadden1978, mcf1} show that when the generator function $G$ satisfies conditions (i)–(iv), the random vector $\boldsymbol{\epsilon} = (\epsilon_1, \ldots, \epsilon_N)$ follows a Multivariate Extreme Value  distribution given by:
\begin{equation}\label{MEV}
F(\epsilon_1, \ldots, \epsilon_N) = \exp\left(-G(e^{-\epsilon_1}, \ldots, e^{-\epsilon_N})\right).
\end{equation}
Moreover, McFadden demonstrates that when $\boldsymbol{\epsilon}$ follows the distribution in \eqref{MEV}, the resulting stochastic choice behavior is consistent with the RUM framework. Importantly, for any deterministic utility vector $\boldsymbol{\theta}$, if we define $\mathbf{y} = (e^{\theta_1}, \ldots, e^{\theta_N})$, the associated social surplus function admits the closed-form expression:
\begin{equation}\label{GEV_Suprlus}
\varphi(\boldsymbol{\theta}) = \log G(\mathbf{y}) + \gamma,
\end{equation}
where $\gamma \approx 0.57721$ is Euler’s constant.

Since the gradient of $\varphi(\boldsymbol{\theta})$ yields the choice probabilities, we obtain:
\begin{eqnarray}\label{GEV_choice}
\nabla_j \varphi(\boldsymbol{\theta}) &=& \frac{\mathbf{y}_j G_j(\mathbf{y})}{\sum_{i=1}^N \mathbf{y}_i G_i(\mathbf{y})} \quad \text{for } j\in A, \\
&=& \frac{e^{\theta_j} G_j(e^{\boldsymbol{\theta}})}{\sum_{i=1}^N e^{\theta_i} G_i(e^{\boldsymbol{\theta}})}. \nonumber
\end{eqnarray}

Equations \eqref{GEV_Suprlus} and \eqref{GEV_choice} illustrate two key properties of the GEV class. First, the choice probability vector takes a logit-like form, ensuring analytical tractability. Second, the closed-form expression for  $\varphi$ facilitates the study of the SSA across a broad class of RUMs.
\subsection{Regret  analysis} We now connect the GEV class with our RUM-ODP framework. Specifically, given the cumulative payoff vector $\boldsymbol{\theta}_t$ and a learning parameter $\eta$, the social surplus function takes the form:
$$
\varphi(\boldsymbol{\theta}_t) = \mathbb{E}\left(\max_{j \in A} \{ \boldsymbol{\theta}_{jt} + \eta \epsilon_{jt+1} \} \right) = \eta \varphi\left( \frac{\boldsymbol{\theta}_t}{\eta} \right).$$

Substituting this into the GEV framework, equations \eqref{GEV_Suprlus} and \eqref{GEV_choice} can be rewritten in the learning context as:
\begin{equation}\label{GEV_Suprlus_Learning}
\varphi(\boldsymbol{\theta}_t) = \eta \left( \log G(e^{\boldsymbol{\theta}t / \eta}) + \gamma \right),
\end{equation}
and
\begin{equation}\label{GEV_choice_Learning}
\nabla_j \varphi(\boldsymbol{\theta}_t) = \frac{e^{\boldsymbol{\theta}{jt}/\eta} G_j(e^{\boldsymbol{\theta}t/\eta})}{\sum_{i=1}^N e^{\boldsymbol{\theta}{it}/\eta} G_i(e^{\boldsymbol{\theta}_t/\eta})} \quad \text{for } j\in A.
\end{equation}

It is important to emphasize that expression \eqref{GEV_choice_Learning} follows from the chain rule, using the fact that for each $j$,
$\frac{\partial \varphi(\boldsymbol{\theta}_t)}{\partial \boldsymbol{\theta}_{jt}} = \eta \cdot \frac{\partial \varphi(\boldsymbol{\theta}_t / \eta)}{\partial \boldsymbol{\theta}_{jt}}.$
This derivation highlights that the GEV model is naturally defined in terms of the scaled utility vector $\boldsymbol{\theta}_t / \eta = \left( \boldsymbol{\theta}_{jt} / \eta \right)_{j \in A}$, consistent with the RUM-ODP learning dynamics.

\begin{example}\label{Logit_Example}Consider the linear aggregator $G(e^{\boldsymbol{\theta}_t/\eta}) = \sum_{i=1}^N e^{\boldsymbol{\theta}_{it}/\eta}$. Under this specification, we recover the classical MNL model. In this case, $\varphi$ becomes:
$$\varphi(\boldsymbol{\theta}_t)
= \eta \varphi\left( \frac{\boldsymbol{\theta}_t}{\eta} \right)
= \eta \log G(e^{\boldsymbol{\theta}_t/\eta}) + \eta \gamma
= \eta \log\left( \sum_{i=1}^N e^{\boldsymbol{\theta}{it}/\eta} \right) + \eta \gamma.$$

The corresponding choice probabilities simplify to the familiar softmax form:
$$\nabla_j \varphi(\boldsymbol{\theta}_t) = \frac{e^{\boldsymbol{\theta}{jt}/\eta}}{\sum_{i=1}^N e^{\boldsymbol{\theta}_{it}/\eta}} \quad \text{for } j\in A.$$

This example illustrates the well-known fact that the MNL model is a special case of the GEV class, where the generator function is the sum of exponentiated payoffs.
\end{example}

By exploiting the structural properties of the GEV class, we can establish that the SSA achieves Hannan consistency. To this end, we rely on the following technical result from \citet[Theorem 3]{Nesterovetal2019}.

\begin{lemma}\label{Result_GEV} Let $\bold{y}_t=(e^{\boldsymbol{\theta}_{1t}/\eta},\ldots,e^{\boldsymbol{\theta}_{Nt}/\eta})\in \RR_+^N$ and let $G$ be  a generator function satisfying the following inequality
	\begin{equation}\label{GEV_condition}
	\sum_{i=1}^{N} \frac{\partial^{2} G(\bold{y}_t)}{\partial \bold{y}_{it}^2} \cdot (\bold{y}_{it})^{ 2} \leq M G(\bold{y}_{1t},\ldots,\bold{y}_{Nt}), \quad\mbox{for all $t=1,\ldots, T$,}
	\end{equation}
for some $M\in \RR_{++}$.
	Then the  social surplus function   $\varphi(\boldsymbol{\theta}_t)=\eta\left(\log G(\bold{y}_t)+ \gamma\right)$ has a  Lipschitz continuous  gradient with constant $L={2M+1\over \eta}.$
		
\end{lemma}

The preceding lemma establishes that the Lipschitz constant L depends on the parameters $M$ and $\eta$. With this result in hand, we can now derive the following regret bound for the GEV class:

\begin{theorem}\label{SS_algorithm_GEV}
Let Assumption \ref{Shocks_Assumption} hold. Furthermore, assume that there exists a generator function $G$ satisfying the inequality in Eq. (\ref{GEV_condition}). Then, under the SSA, the regret satisfies:
\begin{equation}\label{Regret_GEV_SSA}
\textsc{R}^T_{SSA} \leq \eta \log G(\boldsymbol{1}) + \frac{L}{2\eta} T u^2_{max},
\end{equation}
where $L=2M + 1.$ Moreover, by setting $\eta = \sqrt{\frac{L T u^2_{\max}}{2 \log G(\boldsymbol{1})}}$, we obtain:
\begin{equation}\label{Regret_GEV_SS}
\textsc{R}^T_{SSA} \leq u_{max} \sqrt{2 \log G(\boldsymbol{1}) (2M + 1) T}.
\end{equation}
\end{theorem}

Theorem \ref{SS_algorithm_GEV} establishes a regret bound for the SSA when the GEV model serves as the underlying random utility framework. This result shows that, within the RUM-ODP paradigm, the sequence of decisions generated by the SSA is Hannan consistent. To the best of our knowledge, this is the first regret analysis to leverage the full generality of the GEV class. In the following sections, we examine several widely used GEV models that satisfy condition (\ref{GEV_condition}).
\subsection{The GNL model}The most widely used model in the GEV class is the GNL model (\citep{McFadden1978, WenKoppelman2001}). This framework generalizes several other GEV models, including the NL and the OGEV models. Due to its flexibility, the GNL has been successfully applied in various domains such as energy, transportation, housing, telecommunications, and demand estimation.\footnote{For a broader discussion on GEV applications, see \citet{train_2009} and the references therein.}

A key advantage of the GNL model is its ability to incorporate correlation structures among the components of the random shock vector $\epsilon_{t+1}$ in a relatively simple and tractable way.
\smallskip

Let $A$ denote the set of available options, partitioned into $K$ nests labeled $\mathcal{N}_1, \ldots, \mathcal{N}_K$. Let $\mathcal{N}$ be the collection of all such nests. Importantly, the nesting structure permits overlapping nests—that is, an option $i$ can belong to multiple nests simultaneously. For example, an option $i$ may simultaneously be a member of nests $\mathcal{N}_k$, $\mathcal{N}_{k'}$, and $\mathcal{N}_{k''}$.
  \smallskip
  
 For each nest $k = 1, \ldots, K$, let $0 < \lambda_k \leq 1$ denote a nest-specific scale parameter. Economically, $\lambda_k$ reflects the degree of independence among the random shocks $\epsilon_i$ for alternatives belonging to nest $k$. Specifically, the quantity $1 - \lambda_k$ serves as a measure of correlation among those shocks \citep[Ch. 5]{train_2009}. Thus, as $\lambda_k$ increases, the implied correlation within the nest decreases.

\smallskip

Given the cumulative payoff vector $\boldsymbol{\theta}_t$, the generator function $G$ for the GNL model is given by:

\begin{equation}\label{GNL_generating_function}
G(e^{\boldsymbol{\theta}_t/\eta}) = \sum_{k=1}^K \left( \sum_{i=1}^{N} \left( \alpha_{ik} \cdot e^{\boldsymbol{\theta}_{it}/\eta} \right)^{1 / \lambda_k} \right)^{\lambda_k}.
\end{equation}

Here, the parameters $\alpha_{ik} \geq 0$ determine the degree of membership of alternative $i$ in nest $k$. These parameters must satisfy the normalization condition for each alternative $i \in A$:
$$\sum_{k=1}^K \alpha_{ik} = 1.$$

This formulation permits alternatives to belong to multiple nests, with the coefficients $\alpha_{ik}$ representing the degree of association of alternative $i$ with nest $k$.

\smallskip

Using this structure, the set of alternatives assigned to the $k$-th nest can be formally defined as:
$$\mathcal{N}_k = \left\{ i \in A \mid \alpha_{ik} > 0 \right\},$$
where the full choice set is recovered as the union over all nests: $A = \bigcup_{k=1}^K \mathcal{N}_k.$

\smallskip

Under this construction, it can be shown that the generator $G$ given in Eq.~(\ref{GNL_generating_function}) satisfies the required conditions to define a valid GEV model (cf. \citet{WenKoppelman2001}).

To derive the choice probability vector $\bold{x}_{t+1}$ used in the SSA, we note that the GNL model can be seen as operating  through a two-stage decision process.

\smallskip

In the first stage, the DM selects a nest $\mathcal{N}_k $ with probability:
$$
\PP_{kt} = \frac{e^{\bold{v}_{kt}}}{\sum_{\ell=1}^K e^{\bold{v}_{\ell t}}},
$$

where the quantity $\bold{v}_{kt}$ represents the inclusive value (also known as the expected maximum utility) of nest $k$, defined as:
$\bold{v}_{kt} = \lambda_k \log\left( \sum_{i=1}^N \left( \alpha_{ik} \cdot e^{\boldsymbol{\theta}_{it}/\eta} \right)^{1/\lambda_k} \right).$

This inclusive value summarizes the attractiveness of all alternatives within a nest and accounts for both their utilities and the substitution pattern determined by $\lambda_k$.

\smallskip

In the second stage, the DM selects an alternative $i$ from within the previously selected nest $\mathcal{N}_k$ with conditional choice probability:
\[
\PP_{jkt} = \frac{\left( \alpha_{jk} \cdot e^{\boldsymbol{\theta}_{jt}/\eta} \right)^{1/\lambda_k}}{\sum_{i=1}^{N} \left( \alpha_{ik} \cdot e^{\boldsymbol{\theta}_{it}/\eta} \right)^{1/\lambda_k}}.
\]

Combining both stages, the overall probability of selecting alternative $i$ at time $t+1$ under the GNL model is given by:
\[
\bold{x}_{jt+1} = \sum_{k=1}^K \PP_{kt} \cdot \PP_{jkt}, \quad \text{for } j\in A,\ t = 1,\ldots, T-1.
\]
Equivalently, this can be expressed as the gradient of the surplus function: $\nabla_j \varphi(\boldsymbol{\theta}_t) = \bold{x}_{jt+1}$.

\smallskip

Lemma \ref{GNL_regret_lemma} in Appendix \ref{Proofs} establishes two key properties of the GNL model. First, the gradient mapping $\nabla \varphi(\boldsymbol{\theta}_t)$ is Lipschitz continuous with constant
$L = \frac{{2/\min_k \lambda_k - 1}}{\eta}$. Second, the generator function satisfies the bound:
$\log G(\boldsymbol{1}) \leq \log N.$

These properties allow us to derive a regret bound for SSA under the GNL model, as we present next.

\begin{proposition}\label{GNL_regret}
Let $0 < \lambda_k \leq 1$ for all $k = 1, \ldots, K$. In addition, set the learning rate as
$\eta = \sqrt{ \frac{ \left( \frac{2}{\min_k \lambda_k} - 1 \right) T u_{\max}^2 }{ 2 \log N } }$.
Then, under the GNL model, the regret of the SSA satisfies the following bound:
\begin{equation}\label{Bound2_GEV}
\textsc{R}_{SSA}^T \leq u_{max} \sqrt{ 2 \log N \left( \frac{2}{\min_k \lambda_k} - 1 \right) T }.
\end{equation}
\end{proposition}

Proposition \ref{GNL_regret} follows directly from Theorem \ref{SS_algorithm_GEV} and shows that the sequence of decisions generated by the SSA using the GNL model is Hannan consistent under the RUM-ODP framework. This result broadens the applicability of the SSA to more general settings than those allowed by the standard MNL model.

This generalization is made possible by the flexible specification of the generator function $G$, which determines the structure of the choice probabilities. In particular, by modifying the parameters defining the generator $G$, the GNL model can accommodate a wide variety of substitution patterns and correlation structures among alternatives. As described in Appendix \ref{Appendixb}, different specifications of the GNL yield various known models, such as the PCL, the OGEV, and the Principles of Differentiation GEV (PDGEV) model.
\smallskip


\section{ Linking RUM-ODP and FTRL}\label{s4}
The FTRL algorithm models the decision-maker as solving a strictly concave optimization problem that balances cumulative payoffs with a regularization term. This regularizer serves to hedge against unfavorable outcomes or prevent overfitting. The solution at each period of time yields the optimal choice probability distribution based on past payoffs.

In this section, we provide the economic rationale linking the RUM-ODP model to the SSA and FTRL algorithms. Leveraging the convexity of the social surplus function, we derive a strongly convex regularizer $\mathcal{R}(\bold{x})$, defined as the convex conjugate of $\varphi(\boldsymbol{\theta})$. This formulation establishes the equivalence between SSA and FTRL through convex analysis.

We also show the FTRL-generated choice probabilities have a recursive form, extending beyond the MNL’s EWA framework. Finally, the FTRL framework allows the RUM-ODP model to incorporate recency bias effects.
\subsection{Convex conjugate and regularization} From \S\ref{s33}, we know that $\varphi(\boldsymbol{\theta})$ is convex and differentiable. Following \citet{Rockafellar1970}, its convex conjugate is defined as
\[
\varphi^*(\bold{x}) = \sup_{\boldsymbol{\theta} \in \RR^N} \left\{ \langle \boldsymbol{\theta}, \bold{x} \rangle - \varphi(\boldsymbol{\theta}) \right\}.
\]
Using the identity $\varphi(\boldsymbol{\theta}) = \eta \varphi(\boldsymbol{\theta}/\eta)$ and \citet[Thm. 4.14(b)]{Beck2017}, we define the regularization function as $\mathcal{R}(\bold{x}) = \eta \varphi^*(\bold{x}; \eta)$, where $\varphi^*(\bold{x}; \eta)$ is the convex conjugate of the parametrized surplus function $\varphi(\boldsymbol{\theta}/\eta)$. The next result summarizes key properties of $\mathcal{R}(\bold{x})$.

\begin{proposition}\label{R_smooth}
Let Assumptions \ref{Shocks_Assumption} and \ref{Gradient_LL} hold. Then the following statements are true:
\begin{itemize}
    \item[(i)] The function $\mathcal{R}(\mathbf{x})$ is $\frac{\eta}{L}$-strongly convex on $\Delta_N$.
    
    \item[(ii)] The function $\mathcal{R}(\mathbf{x})$ is differentiable for all $\mathbf{x} \in \operatorname{int}(\Delta_N)$.
    
    \item[(iii)] The optimization problem
    \begin{equation}\label{Opt_Program}
        \max_{\mathbf{x} \in \Delta_N} \left\{ \langle \boldsymbol{\theta}, \mathbf{x} \rangle - \mathcal{R}(\mathbf{x}) \right\}
    \end{equation}
   admits a unique solution, which is characterized by
    \[
        \nabla \varphi(\boldsymbol{\theta}) = \arg\max_{\mathbf{x} \in \Delta_N} \left\{ \langle \boldsymbol{\theta}, \mathbf{x} \rangle - \mathcal{R}(\mathbf{x}) \right\}.
    \]
\end{itemize}
\end{proposition}
Part (i) of Proposition \ref{R_smooth} follows from a fundamental equivalence between the Lipschitz continuity of the function $\varphi(\boldsymbol{\theta})$ and the strong convexity of its convex conjugate $\varphi^*(\mathbf{x})$. This relationship is known as the Baillon–Haddad Theorem; see, for example, \citet[12, Section H]{RockafellarWets1997} and \citet{BauschkeCombettes2010}. A formal statement of this result is provided in Lemma \ref{Baillon_Haddad} in Appendix \ref{Proofs}.

Part (ii) establishes the differentiability of $\mathcal{R}(\mathbf{x})$, which is crucial for applying first-order optimality conditions to characterize the optimal solution $\mathbf{x}^*$. This property ensures that standard optimization techniques can be used to solve the problem efficiently.

Part (iii) is a direct consequence of the Fenchel equality, which provides a dual characterization of the optimal solution. An important implication of this result is that, given the cumulative payoff vector $\boldsymbol{\theta}_{t-1}$, the next-period choice probability vector $\mathbf{x}_{t}$ can be equivalently expressed as the unique solution to a strongly concave optimization problem. This perspective offers a useful interpretation of the regularizer $\mathcal{R}(\mathbf{x})$.

Specifically, if $\mathbf{x}_{t} = \nabla \varphi(\boldsymbol{\theta}_{t-1})$, then combining Eq. (\ref{Potential_Econ_concise}) with the Fenchel equality yields the following representation:
\begin{equation}\label{Reg_Weighted}
\mathcal{R}(\mathbf{x}_{t}) = -\eta \sum_{i=1}^N \mathbf{x}_{it} e_{it}(\boldsymbol{\theta}_{t-1}), \quad \text{for } t \geq 1.
\end{equation}

This expression highlights how the distribution of the random perturbations $\epsilon_t$ shapes the form of the regularizer $\mathcal{R}(\bold{x}_t)$. For example, in the MNL model, it is well-known that $e_{jt}(\boldsymbol{\theta}_{t-1}) = -\log \mathbf{x}_{jt}$ for all $j \in A$. Substituting this into Eq. (\ref{Reg_Weighted}) yields the standard entropic regularization: $\mathcal{R}(\mathbf{x}_t) = \eta \sum_{i=1}^N \mathbf{x}_{it} \log \mathbf{x}_{it}$.
\smallskip

The main implication of Proposition \ref{R_smooth} is that the function $\mathcal{R}(\mathbf{x}_t)$ can be interpreted as a regularization term, as discussed in \citet{Hazan2017}.\footnote{From a technical standpoint, \citet{Sandholm2002}, \citet{Abernethyetal2014, Abernethyetal2016}, \citet{GuiyunLiZizhuo2017}, and \citet{Fosgerauetal2019a} establish similar equivalences to (\ref{Opt_Program}). Our proof builds on their arguments and extends their results by establishing the strong convexity of $\mathcal{R}(\bold{x}_t)$.
} This result shows that introducing a stochastic perturbation $\epsilon_t$ into the decision-making process is mathematically equivalent to incorporating a deterministic regularizer 
$\mathcal{R}(\mathbf{x}_t)$. More significantly, this connection provides a foundation for expressing the FTRL algorithm in the following form:

\begin{algorithm}
	\caption{Follow the regularized  leader}\label{alg:euclid2}
	\begin{algorithmic}[1]
		\State Input: $\eta>0$, $\mathcal{R}$, and $\Delta_N$.
		\State Let $\bold{x}_1=\arg\max_{\bold{x}\in\Delta_N}\left\{-\mathcal{R}(\bold{x})\right\}$.
		\State $\bold{for}$ $t=1$ to $T$ \textbf{do}
		\State Produce $\bold{x}_t$.
		\State The environment reveals $\bold{u}_t$.
		\State The DM receives the payoff $\langle\bold{u}_t,\bold{x}_t\rangle$. 
		\State Update $\boldsymbol{\theta}_t=\bold{u}_t+\boldsymbol{\theta}_{t-1}$ and choose
		$$\bold{x}_{t+1}=\arg\max_{\bold{x}\in \Delta_N}\left\{\langle\boldsymbol{\theta}_t,\bold{x}\rangle-\mathcal{R}(\bold{x})\right
		\}.$$
		\State \textbf{end for}
	\end{algorithmic}
\end{algorithm}

\smallskip

We now turn to the central result of this section.

\begin{theorem}\label{Equiv_SSA_FTRL}
Let Assumptions \ref{Shocks_Assumption} and \ref{Gradient_LL} hold. Then the SSA and FTRL algorithms are equivalent.
\end{theorem}

This equivalence relies on Proposition \ref{R_smooth}, which links the regularization function $\mathcal{R}(\mathbf{x}_t)$ to the perturbation-based formulation of SSA. Intuitively, Theorem \ref{Equiv_SSA_FTRL} shows that FTRL and SSA are dual representations of the same learning dynamics. A key technical requirement for this result is that $\varphi(\boldsymbol{\theta}_t)$ has a Lipschitz continuous gradient—a condition that plays a central role in deriving the equivalence. As discussed in Section \ref{s33}, this regularity assumption is satisfied by a broad class of GEV models. Therefore, the FTRL framework offers a powerful lens through which to study no-regret learning algorithms, even in settings that extend well beyond the classical MNL case.

\smallskip


 
\begin{corollary}\label{No_regret_Discrete_choice}Let Assumptions \ref{Shocks_Assumption} and \ref{Gradient_LL} hold. Then the  FTRL Algorithm satisfies the following bound
	$$	\textsc{R}_{FTRL}^T\leq \eta\varphi(\bold{0})+{L\over 2\eta} Tu^2_{\text{max}}.$$
Furthermore, setting $\eta=\sqrt{{LTu^2_{max}\over 2\varphi(\bold{0})}}$ we get  
\begin{equation*}
	\textsc{R}_{FTRL}^T\leq u_{max}\sqrt{2\varphi(\bold{0})LT}.
\end{equation*}	
\end{corollary}

It is important to note that the result in Corollary \ref{No_regret_Discrete_choice} is obtained without requiring knowledge of the explicit functional form of $\mathcal{R}(\mathbf{x})$. While this feature is well-known in the analysis of the FTRL algorithm, the convex structure of the RUM-ODP framework enables a more refined approach: the regret analysis can be carried out using only the information contained in the observed choice probability vector. From a behavioral perspective, this suggests that the FTRL algorithm can be viewed as a learning rule consistent with perturbed random utility models, as discussed in \citet{Sandholm2002}, \citet{Fudenbergetal2015}, and \citet{Fosgerauetal2019a}. More significantly, the result provides an economically grounded optimization interpretation of the FTRL algorithm, further reinforcing its relevance in models of decision-making under uncertainty.

\subsection{Recursive choice}\label{Recursive_Structure}
We now show that under Assumptions \ref{Shocks_Assumption} and \ref{Gradient_LL}, the choice probability vector $\mathbf{x}_{t+1}$ follows a general recursive structure—independent of the specific form of $\mathcal{R}(\textbf{x}_{t+1})$
\smallskip

We begin noticing that the choice probabilities are given by:
	\begin{equation}\label{Grad_logit}
		\bold{x}_{jt+1}={H_{j}(e^{\boldsymbol{\theta}_t/\eta})\over \sum_{i=1}^NH_{i}(e^{ \boldsymbol{\theta}_{t}/\eta})}\quad\mbox{for all $j\in A$, $t\geq 1$},
\end{equation}where the vector-valued function \(H(\cdot)=\left(H_{j}(\cdot)\right)_{j\in A}: \mathbb{R}_{+}^{n} \mapsto \mathbb{R}_{+}^{n}\) is defined
	as the gradient of the exponentiated surplus, i.e.
	\begin{equation}\label{Grad_Exp}
		H\left(e^{\boldsymbol{\theta}_{t}/\eta}\right)=\nabla\left(e^{\varphi\left(\boldsymbol{\theta}_{t}\right) }\right).
	\end{equation}

The expressions in Eqs. (\ref{Grad_logit})–(\ref{Grad_Exp}) were introduced by \citet{Fosgerauetal2019a} to characterize the choice probability vector in discrete choice models that satisfy Assumptions \ref{Shocks_Assumption}. In particular, Eq. (\ref{Grad_logit}) illustrates how the choice probabilities naturally take on a logit-like form. Additionally, \citet[Proposition 2]{Fosgerauetal2019a} establishes that the vector-valued function $H(\cdot)$ is globally invertible, allowing us to define $\Phi(\cdot) \triangleq H^{-1}(\cdot)$. Using this definition, and applying \citet[Proposition 3(ii)]{Fosgerauetal2019a}, we obtain the expression $\mathcal{R}(\mathbf{x}_{t+1}) = \eta \langle \mathbf{x}_{t+1}, \log \Phi(\mathbf{x}_{t+1}) \rangle$. These results together lead to the following proposition:
\begin{proposition}\label{Recursive_Choice_FTRL}
Let Assumptions \ref{Shocks_Assumption} and \ref{Gradient_LL} hold. Then, under the FTRL algorithm, the choice probabilities satisfy the following recursive form:
\begin{equation}\label{Recursive_Prob}
\mathbf{x}_{jt+1} = \frac{H_j\left(e^{\mathbf{u}_t/\eta + \alpha(\mathbf{x}_t)}\right)}{\sum_{i=1}^N H_i\left(e^{\mathbf{u}_t/\eta + \alpha(\mathbf{x}_t)}\right)} \quad \mbox{for all $j \in A, t \geq 1$},
\end{equation}
where $\alpha(\mathbf{x}_t) \triangleq \log \Phi(\mathbf{x}_t)$.
\end{proposition}
Proposition \ref{Recursive_Choice_FTRL} offers a recursive formulation for updating choice probabilities over time. Eq. (\ref{Recursive_Prob}) illustrates how, within the FTRL algorithm, the DM incorporates past information when selecting $\mathbf{x}_{t+1}$, specifically through the term $\alpha(\mathbf{x}_t)$. This term acts as a weight that adjusts the influence of historical performance: alternatives with better past payoffs receive higher weights, while those with worse outcomes are downweighted. For each $j \in A$, the effective payoff is given by $\mathbf{u}_{jt} + \alpha_j(\mathbf{x}_t)$, reflecting accumulated experience.

This formulation also highlights that the structure of the choice probability vector depends on the distributional assumptions about $\epsilon_t$. In particular, the form of $\Phi(\mathbf{x}_t)$—and thus of $\alpha(\mathbf{x}_t)$—is shaped by the underlying distribution of these shocks. As shown in \citet[Proposition 2]{Fosgerauetal2019a}, we have that $-\log \Phi_j(\mathbf{x}_t) = \varphi(\boldsymbol{\theta}_t) - \boldsymbol{\theta}_{jt}$ for all $j \in A$ and $t \geq 1$, linking the regularization structure directly to the distributional features of the model.

To conclude, we note that in the NL model, the regularizer $\mathcal{R}(\mathbf{x}_t)$ admits a closed-form expression, providing further tractability in this particular case.\footnote{Details upon request.}

\subsection{Recency bias}\label{Recency_Bias} Until now, our regret analysis has assumed that the DM assigns equal weight to all past observations. However, substantial empirical evidence suggests that in repeated choice settings, individuals tend to respond more strongly to recent outcomes than to earlier ones—a behavioral pattern known as recency bias (see, e.g., \citet{ErevHaruv2016}, \citet{FudenbergPeysakhovich2014}, \citet{FudenbergLevine2014}).

To incorporate this effect into the RUM-ODP framework, we follow the approach of \citet{RakhlinSridharan2013} by introducing a sequence of functions $\boldsymbol{\beta}_t: \mathcal{U}^{t-1} \mapsto \mathcal{U}$, for $t = 1, \ldots, T$, which generates a predictable sequence reflecting the DM’s evolving beliefs or forecasts
\begin{equation}\label{Predictable_Sequence}
 \boldsymbol{\beta}_1(\boldsymbol{0}),\boldsymbol{\beta}_{2}(\bold{u}_1),\ldots, \boldsymbol{\beta}_T(\bold{u}_1,\ldots,\bold{u}_{T-1}).
\end{equation}

The sequence defined in (\ref{Predictable_Sequence}) serves as a way to incorporate prior beliefs or knowledge about the payoff sequence $\mathbf{u}_1, \ldots, \mathbf{u}_T$. By choosing appropriate summary statistics, this sequence allows us to capture both the presence and magnitude of recency bias. To account for this behavioral feature, we modify the FTRL algorithm by incorporating the predictable sequence into the update rule as follows:\footnote{For notational simplicity, we write $\boldsymbol{\beta}_t(\mathbf{u}_1, \ldots, \mathbf{u}_{t-1}) \triangleq \boldsymbol{\beta}_t$ for $t = 1, \ldots, T.$}
\begin{equation}\label{OFTRL_Eq1}
\bold{x}_{t+1}=\begin{cases} \arg\max_{\bold{x}\in \Delta_N}\{-\mathcal{R}(\bold{x})\} &\mbox{for } t = 0 \\ 
\arg\max_{\bold{x}\in \Delta_N}\{\langle\boldsymbol{\theta}_t+\boldsymbol{\beta}_t,\bold{x}\rangle-\mathcal{R}(\bold{x})\} & \mbox{for } t=1,\ldots, T. \end{cases}.
\end{equation}

Expression (\ref{OFTRL_Eq1}) makes explicit that by adding the term $\boldsymbol{\beta}_t$, we obtain a variant of the FTRL algorithm that incorporates recency bias. Following \citet{RakhlinSridharan2013} and \citet{NIPS2015_5763}, we refer to the resulting algorithm as Optimistic FTRL (OFTRL). We specifically examine the case of $S$-step recency bias:

\begin{definition}\label{Def_Recency_Bias}
In the OFTRL algorithm, we say that the DM exhibits $S$-step recency bias if
$\boldsymbol{\beta}_t = \frac{1}{S} \sum_{\tau = t-S}^{t-1} \bold{u}_\tau$
for  $t = 1, \ldots, T.$
\end{definition}

The following proposition establishes that, under $S$-step recency bias, the OFTRL algorithm remains Hannan consistent.
\begin{proposition}\label{OFTRL_Regret}
Let Assumptions \ref{Shocks_Assumption} and \ref{Gradient_LL} hold. Additionally, suppose the utility sequence satisfies the following bounded variation condition:
\begin{equation}\label{Bound_Dual}
\|\bold{u}_t - \bold{u}_{t-1}\|_\ast \leq B \quad \text{for all } t = 1, \ldots, T.
\end{equation}
Then, under $S$-step recency bias and with learning rate
$\eta = \sqrt{\frac{L T S^2 B^2}{2\varphi(\bold{0})}}$,
the OFTRL algorithm is Hannan consistent with regret bounded as:
\[
\textsc{R}_{\textsc{OFTRL}}^T \leq SB \sqrt{2LT \varphi(\bold{0})}.
\]
\end{proposition}

To understand the intuition behind Proposition \ref{OFTRL_Regret}, we refer to Lemma \ref{Bound_OFTRL} in Appendix \ref{Appendixb}, which shows:
\begin{equation}\label{OFRTL_bound_general}
\textsc{R}_{\textsc{OFTRL}}^T \leq \eta \varphi(\bold{0}) + \frac{L}{2\eta} \sum_{t=1}^T \|\bold{u}_t - \boldsymbol{\beta}t\|_*^2.
\end{equation}

This bound implies that regret is low when the prediction sequence $\boldsymbol{\beta}_t$ approximates $\bold{u}_t$ well. In the case of $S$-step recency bias, $\boldsymbol{\beta}_t$ is a moving average of recent utilities. When the utility sequence varies slowly, this leads to accurate predictions and thus low regret. This fact combined with Lemma \ref{Cases_regret} in \ref{Appendixb} completes the proof of Proposition \ref{OFTRL_Regret}.

\smallskip

Technically, the proof of Proposition \ref{OFTRL_Regret} adapts ideas from \citet{RakhlinSridharan2013} and \citet{NIPS2015_5763}, but differs in its use of the convex structure of the RUM. This gives an economic interpretation to the performance of OFTRL under recency bias.

\smallskip

Finally, using Theorem \ref{Equiv_SSA_FTRL}, we derive an equivalent optimistic SSA formulation. Under recency bias, the social surplus becomes:
$$\varphi(\boldsymbol{\theta}_t + \boldsymbol{\beta}_{t+1}) = \mathbb{E} \left[ \max_{j \in A} \{ \boldsymbol{\theta}_{jt} + \boldsymbol{\beta}_{jt+1} + \epsilon_{jt+1} \} \right],$$
with the corresponding choice rule:
$$\bold{x}_{t+1} = \nabla \varphi(\boldsymbol{\theta}_t + \boldsymbol{\beta}_{t+1}) = \arg\max_{\bold{x} \in \Delta_N} \left\{ \langle \boldsymbol{\theta}_t + \boldsymbol{\beta}_{t+1}, \bold{x} \rangle - \mathcal{R}(\bold{x}) \right\}.$$

This highlights a natural link between recency bias, optimism, and RUM-based decision-making.

\section{No-regret learning in games}\label{S5_Games}
In this section, we apply the RUM-ODP model to study no-regret learning in games. We consider a static game $\mathcal{G}$ with a set $\mathcal{P}$ of $P$ players.\footnote{This section closely follows the exposition in \cite{NIPS2015_5763}.} Each player $p$ has a finite strategy set $S_p$ and a utility function
$\textbf{u}_p : S_1 \times \cdots \times S_P \rightarrow [0,1]$,
which assigns a utility $u_p(\bold{s})$ to each strategy profile $\bold{s} = (s_1, \ldots, s_P)$. We assume each $S_p$ has cardinality $N$, i.e., $|S_p| = N$ for all $p\in \P$.

Let $\mathbf{x} = (\mathbf{x}_1, \ldots, \mathbf{x}_P)$ denote a mixed strategy profile, where $\mathbf{x}_p \in \Delta_N$ and $\mathbf{x}_{pk}$ is the probability assigned to pure strategy $k \in S_p$ by player $p$.\footnote{In this context, we use the notation $\Delta(S_p) \triangleq \Delta_N$ for the mixed strategy simplex.} The set of all mixed strategy profiles is $\Delta_N^P \triangleq \prod_{p \in \mathcal{P}} \Delta_N$. Finally, define the expected utility of player $j$ under profile $\mathbf{x}$ as: $U_p(\mathbf{x}) = \mathbb{E}_{\bold{s} \sim \bold{x}} \left[ \textbf{u}_p(\bold{s}) \right].$
\smallskip

We consider a setting where the game $\mathcal{G}$ is played repeatedly over $T$ time steps. We denote this repeated game by $\mathcal{G}^T$. At each time step $t$, each player $p$ selects a mixed strategy $\bold{x}_{pt} \in \Delta_N$. At the end of each iteration, player $j$ observes the expected utility they would have obtained had they played any pure strategy $k \in S_p$, given the mixed strategies of the other players. Formally, let: $\textbf{u}_{pkt} = \mathbb{E}_{\bold{s}_{-p} \sim \bold{x}_{-pt}} \left[ \textbf{u}_p(k, \bold{s}_{-p}) \right],$
where $\bold{s}_{-p}$ denotes the strategies of all players other than player $p$, and $\bold{x}_{-pt}$ is the profile of their mixed strategies at time $t$.
Let $\bold{u}_{pt} = (\textbf{u}_{pkt})_{k \in S_p}$ denote the feedback vector observed by player $p$ at time $t$. The expected utility of player $p$ at iteration $t$ is then given by:
$\left\langle \bold{x}_{pt}, \bold{u}_{pt} \right\rangle.$
\smallskip

To model no-regret learning, we assume that each player updates their mixed strategy $\bold{x}_{pt+1}$ using the SSA. Specifically, player $p$’s social surplus function is defined as:
$\varphi_p(\boldsymbol{\theta}_{pt}) \triangleq \mathbb{E} \left[\max_{k\in S_p} \left\{ \boldsymbol{\theta}_{pkt} + \eta_p \epsilon_{pkt+1} \right\} \right],$ where the expectation is taken over the random vector $\boldsymbol{\epsilon}_{pt+1}$, and $\eta_p > 0$ is a player-specific learning rate.

\smallskip

In this strategic setting, the SSA dynamics for each player $p$ are defined as follows. The initial strategy is set to:
$\bold{x}_{p0} = \nabla \varphi_p(\boldsymbol{0}),$
and for each time step $t = 1, \ldots, T$, the update rule is:
$$\bold{x}_{pt+1} = \nabla \varphi_p(\boldsymbol{\theta}_{pt}) \quad\mbox{for all $p \in \mathcal{P}$},$$
where the cumulative utility vector is defined as:
$\boldsymbol{\theta}_{pt} \triangleq \sum_{l=1}^t \bold{u}_{pl}.$

\smallskip

In the repeated game $\mathcal{G}^T$, the regret of player $p \in \mathcal{P}$ after $T$ periods is defined as the maximum cumulative gain they could have achieved by deviating to the best fixed strategy in hindsight:
\[
\textsc{R}_{\text{SSA}_p}^T \triangleq \max_{\bold{x}_p^* \in \Delta_N} \sum_{t=1}^{T} \left\langle \bold{x}_p^* - \bold{x}_{pt}, \bold{u}_{pt} \right\rangle.
\]

\smallskip

It is straightforward to show that under  Assumptions \ref{Shocks_Assumption} and \ref{Gradient_LL}, we can apply  Theorem \ref{SS_Surplus_algorithm_regret} to bound $\textsc{R}_{\text{SSA}_p}^T$. In particular,  in the repeated  game $\G^T,$ setting  $\eta_p=\sqrt{{L_pT\over 2\varphi_p(\bold{0})}}$, we obtain\footnote{We note that in setting  $\eta_p=\sqrt{{L_T\over 2\varphi_p(\bold{0})}}$ we have used the fact that $u_{max}=1$ for all player $p\in \P.$ In addition,  the parameter $L_p$ corresponds to the Lipschitz constant associated to  player $p$'s social surplus function $\varphi_p$.}
\begin{equation*}
	\textsc{R}_{\text{SSA}_p}^T\leq \sqrt{2\varphi_p(\bold{0})L_pT}\quad\mbox{for all $p\in \P$}.
\end{equation*}	

The main implication of using the RUM-ODP model combined with the SSA  is that we can bound the regret associated with each player using a large class of discrete choice models. For instance, we can consider cases where some players may use an MNL model, and others can use GNL. In general, our approach is flexible enough to accommodate players using different discrete choice models to compute $\nabla\varphi_p(\boldsymbol{\theta}_{pt})=\textbf{x}_{pt+1}$ for each player $p\in\P.$
\smallskip

A key advantage of using SSA to study no-regret learning in normal-form games is that it avoids the need to specify a regularization term $\mathcal{R}_p(\bold{x}_{pt})$—unlike much of the existing literature on regularized learning.\footnote{See, e.g., \cite{NIPS2015_5763} and \cite{Sandholm_Mertikopoulos_2016}.} This makes SSA applicable even when $\mathcal{R}_p$ lacks a closed-form expression. Moreover, as shown in \S\ref{Recency_Bias}, SSA can naturally incorporate recency bias, broadening its relevance to more realistic learning behaviors. 
\subsection{ Coarse Correlated Equilibrium}Using the RUM-ODP model to study no-regret learning in games directly expands the class of choice models that can approximate coarse correlated equilibrium (CCE).
\begin{definition}[Coarse Correlated Equilibrium ]
	A distribution $\sigma$ on the set $S_{1} \times \cdots \times S_{P}$ of outcomes of  the  game  $\G$ is a coarse correlated equilibrium  if for every agent $p \in\P$ and every unilateral deviation $s_{p}^{\prime} \in S_{p}$,
	\begin{equation}\label{CCE_Condition}
		\mathbb{E}_{\mathbf{s} \sim \sigma}\left[\bold{u}_{p}(\mathbf{s})\right] \geq \mathbb{E}_{\mathbf{s} \sim \sigma}\left[\bold{u}_{p}\left(s_{p}{ }^{\prime}, \mathbf{s}_{-p}\right)\right] .
	\end{equation}
\end{definition}
Condition (\ref{CCE_Condition}) matches that of a mixed strategy Nash equilibrium, but without requiring $\sigma$ to be a product distribution. Intuitively, it captures a setting where player $i$ considers deviating to $s_i’$ knowing only the joint distribution $\sigma$, not their actual action $s_i$. Thus, a CCE protects only against unconditional unilateral deviations, unlike a correlated equilibrium  (CE), which accounts for deviations conditioned on $s_i$ (\cite{Aumann197467}). Since every CE is also a CCE, a CCE always exists and is computationally tractable. More importantly, no-regret dynamics are known to converge to the set of CCEs. The following result extends Proposition 17.9 in \cite{Roughgarden2016} to the RUM-ODP model; the proof is omitted.

\begin{proposition}\label{CCE_Result} Let Assumptions  \ref{Shocks_Assumption} and \ref{Gradient_LL} hold.  Suppose that at periods $t=1,\ldots, T$,  players choose their strategy $\bold{x}_{pt}$ according to the SSA. Let $\sigma_{t}=\prod_{p=1}^{P} \bold{x}_{pt}$ denote the outcome distribution at iteration $t$ and $\sigma=\frac{1}{T} \sum_{t=1}^{T} \sigma_{t}$ the time-averaged history of these distributions. Then $\sigma$ is an approximate coarse correlated equilibrium, in the sense that
	$$
	\mathbb{E}_{\mathbf{s} \sim \sigma}\left[\bold{u}_{p}(\mathbf{s})\right] \geq \mathbb{E}_{\mathbf{s} \sim \sigma}\left[\bold{u}_{p}\left(s_{p}^{\prime}, \mathbf{s}_{-p}\right)\right]-\delta.
	$$
	for every agent $i$ and unilateral deviation $s_{i}^{\prime}$ where $\delta\triangleq \max_{p\in \P}\{\textsc{R}_{\text{SSA}_p}^T\}$.
\end{proposition}

Proposition \ref{CCE_Result} extends a standard no-regret learning result by linking the regret bound $\delta$ to the social surplus function $\varphi$ and Lipschitz constant $L$, showing how the RUM structure influences convergence to CCE.

Unlike traditional approaches that rely on fixed regularization functions—typically tied to the MNL—SSA accommodates a broad class of RUMs, even when the regularizer lacks a closed form. This generality allows for greater flexibility in modeling learning behavior in games.

Although related to potential-based dynamics, our use of the social surplus function differs in key ways. It avoids restrictive assumptions like bounded shocks or specific gradient conditions and offers a clear economic interpretation based on cumulative payoffs and preference shocks. 

\section{Related literature}\label{s5}
Regret theory, introduced by \citet{Bell1982}, \citet{LoomesSugden1982}, and \citet{FISHBURN198231}, offers an alternative to expected utility by capturing the desire to avoid outcomes that may, in hindsight, feel like poor decisions—often resulting in non-transitive preferences (\citet{BleichrodtWaakker2015}). More recent axiomatizations are provided in \citet{Sarver2008} and \citet{HAYASHI2008242}. Building on this, we analyze regret algorithmically within the RUM-ODP framework, aligning more closely with the algorithmic game theory literature (\citet{Roughgarden2016}).

We also contribute to the no-regret learning literature originating with \citet{Hahn1957} and extended by \citet{LittlestoneWarmuth1994}, \citet{FudenbergLevine1995}, \citet{FreundSchapire1997}, and \citet{FosterVohra1997}. More recent work, such as \citet{NIPS2015_5763}, investigates fast convergence in online learning via FTRL. Our approach differs by (i) introducing the SSA for analyzing RUM-ODP models through discrete choice theory, and (ii) generalizing the EWA and regret framework to a broader class of RUMs. Related studies such as \citet{BianchiLugosi2003}, \citet{HART_JET_200126}, and \citet{HART_GEB_2003375} rely on potential functions in no-regret learning. While SSA is also a potential-based method, it avoids their assumptions of additivity and vanishing gradients, which are incompatible with RUMs, thereby broadening applicability.

We further model the RUM-ODP as a two-player game, connecting with \citet{gualdani2020identification}, who study Bayes Correlated Equilibrium in static games with incomplete information. In contrast, our approach (i) focuses on repeated decisions, (ii) emphasizes Hannan consistency, and (iii) operates within discrete choice settings. Similarly, while \citet{Magnolfi2021} address identification in Bayesian games, they do not examine RUMs.

Our work contributes to the online convex optimization (OCO) literature (\citet{Shalev-Shwartz2012}; \citet{Hazan2017}), particularly extending insights from \citet{Abernethyetal2016}. We distinguish our contribution by: (i) applying SSA to derive concrete RUM examples such as NL, GNL, and GEV; (ii) uncovering a recursive EWA-like structure for RUMs; and (iii) deriving a closed-form regularizer for the NL model, which was not addressed in their work.

We also relate to the Rational Inattention (RI) literature (\citet{CaplinDean2015}; \citet{Matejka2015}), where stochastic choice arises from information acquisition costs. Unlike these static models, we consider repeated choices without priors or belief updates, and adopt regret—not utility maximization—as the performance criterion.

\citet{WEBB2019} and \citet{cerreiavioglio2021multinomial} connect RUMs with decision times and neural mechanisms. In contrast, we focus on learning through repeated choice, leveraging tools from online optimization and regret minimization rather than neuro-inspired modeling.

Finally, \citet{Fudenbergetal2015} interpret stochastic choice as utility maximization under regularization. While we also employ regularization, our emphasis is on dynamic learning and performance guarantees via regret, not on axiomatic foundations. We further draw on \citet{Nesterovetal2019}, who link social surplus to proxy functions, though their work does not address no-regret dynamics.

\section{Final Remarks }\label{s6}
This paper introduced the RUM-ODP model, linking RUMs with gradient-based learning to study no-regret behavior in uncertain environments. We develop the SSA framework, which generalizes online decision-making across a broad class of discrete choice models. Our analysis offers a behavioral interpretation of the FTRL algorithm and extends the widely EWA approach far beyond the MNL model. As an application, we discussed how how our approach is useful to study no-regret dynamics in games.

\bibliographystyle{abbrvnat}
\bibliography{References_RUM_OCO}


\appendix
\section{Proofs}\label{Proofs}

We begin stating two technical results that will be used throughout this appendix.

\begin{lemma}[Baillon-Haddad Theorem]\label{Baillon_Haddad} The following statements are equivalent
	\begin{itemize}
		\item[i)] $h: \mathbb{E} \rightarrow \mathbb{R}$ is convex  and differentiable with  gradient $\nabla h$ which is Lipschitz continuous with
		respect to $\|\cdot\|_{\mathbb{E}}$ with constant $L>0$.
		\item[ii)]The convex conjugate $h^{*}: \mathbb{E}^{*} \rightarrow(-\infty, \infty]$ is ${1\over L}$-strongly convex with respect to  the dual norm $\|\cdot\|_{\mathbb{E}^*}^{*}$.
	\end{itemize}
\end{lemma}
\proof\citet[ Thm. 12, Section H]{RockafellarWets1997}.\footnote{We remark that in this theorem $\mathbb{E}^{*}$ denotes the dual space of $\mathbb{E}$ and $\|\cdot\|^*_{\mathbb{E}^*}$ denotes its corresponding dual norm.} \eproof

\begin{lemma}\label{Convexity_smoothness}Let Assumption \ref{Shocks_Assumption}  hold. Then $\mathcal{R}$ is differentiable.
\end{lemma}\proof The proof follows from a direct application of \citet[Thm. 5]{SorensenForsgerau2020} or \citet[Thm. 2]{galichon2021cupids}.\eproof

 \subsection{Proof of Proposition \ref{SS_Equiv}} Note that  by definition  $\varphi(\boldsymbol{\theta}_t)=\EE[\tilde{\varphi}(\boldsymbol{\theta}_t+\eta\epsilon_{t+1})]$. Then combining (\ref{FTPL3}) with \citet[{Prop. 2.3}]{Bertsekas1973} it follows that 
\begin{eqnarray}
\tilde{\bold{x}}_{t+1}&\in & \partial\tilde{\varphi}(\boldsymbol{\theta}_t+\eta\epsilon_{t+1}),\nonumber\\
\EE[\tilde{\bold{x}}_{t+1}]&\in &\EE[\partial\tilde{\varphi}(\boldsymbol{\theta}_{t}+\eta\epsilon_{t+1})],\nonumber\\
&=& \partial\EE[\tilde{\varphi}(\boldsymbol{\theta}_t+\eta\epsilon_{t+1})],\nonumber\\
&=&\nabla\EE[\tilde{\varphi}(\boldsymbol{\theta}_t+\eta\epsilon_{t+1})]=\nabla\varphi(\boldsymbol{\theta}_t).\nonumber
\end{eqnarray}\eproof

\subsection{Proof of Lemma \ref{Gradient_Lipschitz}} In proving this lemma we use the fact  that  $$\varphi(\boldsymbol{\theta})=\EE\left[\max_{j\in A}\{\boldsymbol{\theta}_{j}+\eta\epsilon_j\}\right]=\eta\EE\left[\max_{j\in A}\{\boldsymbol{\theta}_{j}/\eta+\epsilon_j\}\right]=\eta\varphi(\boldsymbol{\theta}/\eta).$$ 
Thus, it is easy to see that 
$\nabla\varphi(\boldsymbol{\theta})=\nabla\varphi(\boldsymbol{\theta}/\eta)$. In addition, simple algebra shows that  ${\partial^2\varphi(\boldsymbol{\theta})\over \partial \boldsymbol{\theta}_j\partial \boldsymbol{\theta}_i}={1\over \eta }{\partial^2\varphi(\boldsymbol{\theta}/\eta)\over \partial \boldsymbol{\theta}_j\partial \boldsymbol{\theta}_i},$ for all $i,j=1,\ldots, N.$ Under this equivalence we note that 
the condition in Assumption \ref{Gradient_LL} can be rewritten as $2Tr(\varphi(\boldsymbol{\theta}/\eta))\leq L$.
\smallskip

Define the function $f(t)=\nabla\varphi(\boldsymbol{\theta}_1/\eta+t(\boldsymbol{\theta}_2/\eta-\boldsymbol{\theta}_1/\eta))$  with $f^\prime(t)=\langle\nabla^2\varphi(\boldsymbol{\theta}_1/\eta+t(\boldsymbol{\theta}_2/\eta-\boldsymbol{\theta}_1/\eta)),\boldsymbol{\theta}_2/\eta-
\boldsymbol{\theta}_1/\eta\rangle$. Noticing that 
\begin{eqnarray}
\nabla\varphi(\boldsymbol{\theta}_2/\eta)-\nabla\varphi(\boldsymbol{\theta}_1/\eta)=f(1)-f(0)&=&\int_{0}^1f^\prime(t)dt\nonumber\\
&=&\int_{0}^1\nabla^2\varphi(\boldsymbol{\theta}_1/\eta+t(\boldsymbol{\theta}_2/\eta-\boldsymbol{\theta}_1/\eta))(\boldsymbol{\theta}_2/\eta-\boldsymbol{\theta}_1/\eta)dt.\nonumber\\
\|\nabla\varphi(\boldsymbol{\theta}_2/\eta)-\nabla\varphi(\boldsymbol{\theta}_1/\eta)\|_1&\leq& \int_{0}^1\|\nabla^2\varphi(\boldsymbol{\theta}_1/\eta+t(\boldsymbol{\theta}_2/\eta-\boldsymbol{\theta}_1/\eta))(\boldsymbol{\theta}_2/\eta-
\boldsymbol{\theta}_1/\eta)\|_1dt\nonumber\\
&\leq &\int_{0}^1\|\nabla^2\varphi(\boldsymbol{\theta}_1/\eta+t(\boldsymbol{\theta}_2/\eta-\boldsymbol{\theta}_1/\eta))\|_{\infty,1}\|\boldsymbol{\theta}_2/\eta-\boldsymbol{\theta}_1/\eta\|_1\nonumber
\end{eqnarray}

To complete the proof we stress two properties of the Hessian. First, each row (or columns) of $\nabla^2\varphi(\boldsymbol{\theta}_t/\eta)$ sums up to $0$. To see this we note that $\sum_{i=1}^N\nabla_i\varphi(\boldsymbol{\theta}_t/\eta)=1$. Then simple differentiation yields $\sum_{j=1}^N{1\over \eta}\nabla_{ij}\varphi(\boldsymbol{\theta}_t/\eta)=0$ for all $i=1,\ldots, N.$ Second,  it is well known that  the off-diagonal elements of ${1\over \eta}\nabla^2\varphi(\boldsymbol{\theta}_t/\eta)$ are nonnegative (\citet[Ch. 5]{mcf1}). To see why this is true, we recall that for alternative $i$ the choice probability is given by: $\nabla_i\varphi(\boldsymbol{\theta}_t/\eta)=\PP(i=\arg\max_{j\in A}\{\boldsymbol{\theta}_{jt}+\eta\epsilon_{jt}\})$. Then increasing the terms $\boldsymbol{\theta}_{jt}$ for $j\neq i$  cannot increase the probability of choosing $i$, which is formalized as ${1\over \eta}\nabla_{ij}\varphi(\boldsymbol{\theta}_t/\eta)\leq 0$.

Now, using previous observation, we have that for a convex combination  $\tilde{\boldsymbol{\theta}}/ \eta=\boldsymbol{\theta}_1/ \eta+t(\boldsymbol{\theta}_2/ \eta-\boldsymbol{\theta}_1/ \eta)$ we have
\begin{eqnarray}
{1\over \eta}\|\nabla^2\varphi(\tilde{\boldsymbol{\theta}}/\eta)\|_{\infty,1}&=&{1\over \eta}\max_{\|\bold{v}\|\leq 1}\{\|\nabla^2\varphi(\tilde{\boldsymbol{\theta}}/\eta)\bold{v}\|_1\}\nonumber\\
&\leq& {1\over \eta}\sum_{i=1}^N\sum_{j=1}^N|\nabla_{ij}^2\varphi(\tilde{\boldsymbol{\theta}})|\nonumber\\
&=& {1\over \eta}2Tr(\nabla^2\varphi(\tilde{\boldsymbol{\theta}}))\leq {L\over \eta},\nonumber
\end{eqnarray}
where the last inequality follows from Assumption \ref{Gradient_LL}. 
Plugging  in, we arrive to the conclusion
$$\|\nabla\varphi(\boldsymbol{\theta}_2/\eta)-\nabla\varphi(\boldsymbol{\theta}_1/\eta)\|_1\leq {L }\|\boldsymbol{\theta}_2/\eta-\boldsymbol{\theta}_1/\eta\|_1\quad\forall \boldsymbol{\theta}_1,\boldsymbol{\theta}_2.$$

Finally, using the fact $\nabla\varphi(\boldsymbol{\theta})=\nabla\varphi(\boldsymbol{\theta}/\eta)$
we get
$$\|\nabla\varphi(\boldsymbol{\theta}_2)-\nabla\varphi(\boldsymbol{\theta}_1)\|_1\leq {L\over \eta }\|\boldsymbol{\theta}_2-\boldsymbol{\theta}_1\|_1\quad\forall \boldsymbol{\theta}_1,\boldsymbol{\theta}_2.$$
\eproof
\begin{lemma}\label{Positive_R_Bound_D}Let Assumption \ref{Shocks_Assumption} hold. Then	$\mathcal{R}(\bold{x})\leq 0$ for all $\bold{x}\in \Delta_N$.
	
\end{lemma}
\proof  First, we note that the Fenchel equality implies $\mathcal{R}(\bold{x})=\langle \boldsymbol{\theta},
\bold{x}\rangle-\varphi(\boldsymbol{\theta})=\langle \boldsymbol{\theta},
\bold{x}\rangle-\eta\varphi(\boldsymbol{\theta}/\eta)$ with $\bold{x}=\nabla\varphi(\boldsymbol{\theta})=\nabla\varphi(\boldsymbol{\theta}/\eta)$. We recall  that $\varphi(\boldsymbol{\theta})=\EE[\max_{j\in A}\{\boldsymbol{\theta}_j+\eta\epsilon_j\}]$. Given that  $\max\{\cdot
\}$ is a convex function, by Jensen's inequality we get $\max_{j\in A}\EE[\boldsymbol{\theta}_j+\eta\epsilon_j]\leq \EE[\max_{j\in A}\{\boldsymbol{\theta}_j+\eta\epsilon_j\}].$
Then it follows that 
\begin{eqnarray}
\mathcal{R}(\bold{x})&=& \langle \boldsymbol{\theta},\bold{x}\rangle-\varphi(\boldsymbol{\theta}),\nonumber\\
&\leq&\langle \boldsymbol{\theta},\bold{x}\rangle-\max_{j\in A}\EE[\boldsymbol{\theta}_j+\eta\epsilon_j]\nonumber\\
&=& \langle \boldsymbol{\theta},\bold{x}\rangle-\max_{j\in A}\boldsymbol{\theta}_j,\nonumber\\
\mathcal{R}(\bold{x})&\leq& 0\nonumber
\end{eqnarray}
where the last inequality follows from the fact  that $\bold{x}\in \Delta_N$.\eproof

\begin{lemma}\label{Bregman_Bound}Let Assumptions \ref{Shocks_Assumption} and \ref{Gradient_LL} hold. Then
	$$D_{\varphi}(\boldsymbol{\theta}_t||\boldsymbol{\theta}_{t-1})\leq {L\over 2\eta}u^2_{max}.$$
\end{lemma}
\proof 
Using a second order Taylor expansion of $\varphi(\boldsymbol{\theta}_t)$ we get:

\begin{eqnarray}
\varphi(\boldsymbol{\theta}_{t+1})&=&\varphi(\boldsymbol{\theta}_t)+\langle\nabla\varphi(\boldsymbol{\tilde{\theta}}),\bold{u}_t\rangle+{1\over 2}\langle\bold{u}_t,\nabla^2\varphi(\tilde{\boldsymbol{\theta}})\bold{u}_t\rangle,\nonumber\\
\varphi(\boldsymbol{\theta}_{t+1})-\varphi(\boldsymbol{\theta}_t)-\langle\nabla\varphi(\boldsymbol{\tilde{\theta}}),\bold{u}_t\rangle&=&{1\over 2}\langle\bold{u}_t,\nabla^2\varphi(\tilde{\boldsymbol{\theta}})\bold{u}_t\rangle,\nonumber\\
D_\varphi(\boldsymbol{\theta}_{t+1}||\boldsymbol{\theta}_t)&=&{1\over 2}\langle\bold{u}_t,\nabla^2\varphi(\tilde{\boldsymbol{\theta}})\bold{u}_t\rangle,\label{Bregman_1}
\end{eqnarray}
where $\tilde{\boldsymbol{\theta}}$ is some convex combination of $\boldsymbol{\theta}_{t+1}$ and $\boldsymbol{\theta}_t$. From Eq.(\ref{Bregman_1})  it follows that
\begin{equation}\label{Bregman2}
D_\varphi(\boldsymbol{\theta}_{t+1}||\boldsymbol{\theta}_t)\leq {1\over 2}\|\nabla_{ij}^2\varphi(\tilde{\boldsymbol{\theta}})\|_{\infty,1}\|\bold{u}_t\|_{\infty}^2.\end{equation}

Noting that $$\|\nabla^2\varphi(\tilde{\boldsymbol{\theta}})\|_{\infty,1}=\max_{\|\bold{v}\|\leq 1}\{\|\nabla^2\varphi(\tilde{\boldsymbol{\theta}})\bold{v}\|_1\}\leq \sum_{i=1}^N\sum_{j=1}^N|\nabla_{ij}^2\varphi(\tilde{\boldsymbol{\theta}})|=2Tr(\nabla^2\varphi(\tilde{\boldsymbol{\theta}}))\leq {L\over \eta},$$
where the last inequality follows from Assumption \ref{Gradient_LL}.

 Plugging the previous bound in (\ref{Bregman2}) combined with $\|\bold{u}_t\|^2_{\infty}\leq u_{max}^2$    we find that                                                                                                                                                                                                                                                                                                                                                                                                                                                                                                                                                                                                                                                                                                                                                                                                                                                                                                                                                                                                                                                                                                                                                                                                                                                                                                                                                                                                                                                                                                                                                                                                                                                                                                                                                                                                                                                                                                                                                                                                                                                                                                                                                                                                                                                                                                                                                                                                                                                                                                                                                                                                                                                                                                                                                                                                                                                                                                                                                                                                                                                                                                                                                                                                                                                                                                                                                                              
$$D_\varphi(\boldsymbol{\theta}_{t+1}||\boldsymbol{\theta}_t)\leq{L\over 2\eta}u_{max}^2.$$\eproof

\begin{lemma}\label{Bound_Regret_FTRL}Let Assumptions \ref{Shocks_Assumption} and \ref{Gradient_LL} hold. Then in the SSA 
	\begin{equation}\label{FRTL_Bounded}
	\textsc{R}_{SSA}^T\leq \eta\varphi(\bold{0})+{L\over 2\eta}Tu^2_{max}.
	\end{equation}
\end{lemma}
\proof The proof of this lemma exploits the convex duality structure of the RUM-ODP model. 
By the Fenchel-Young inequality we know that 

$$\forall\bold{x}\in\Delta_N:\quad\mathcal{R}(\bold{x})\geq \langle\boldsymbol{\theta}_T,\bold{x}\rangle-\varphi(\boldsymbol{\theta}_T),$$
where the equality holds when $\bold{x}$ maximizes $\langle\boldsymbol{\theta}_T,\bold{x}\rangle-\mathcal{R}(\bold{x})$.

The Fenchel-Young inequality implies
$$\mathcal{R}(\bold{x})-\langle\boldsymbol{\theta}_T,\bold{x}\rangle\geq -\varphi(\boldsymbol{\theta}_T).$$

Noting that $-\varphi(\boldsymbol{\theta}_T)$ can be equivalently expressed as:
$$-\varphi(\boldsymbol{\theta}_T)=-\varphi(\boldsymbol{\bold{0}})-\sum_{t=1}^T\left(\varphi(\boldsymbol{\theta}_{t})-\varphi(\boldsymbol{\theta}_{t-1})\right).$$

From the definition of Bregman divergence combined with  $\bold{x}_t=\nabla\varphi(\boldsymbol{\theta}_{t-1})$, it follows that 
$$\sum_{t=1}^T\left(\varphi(\boldsymbol{\theta}_{t})-\varphi(\boldsymbol{\theta}_{t-1})\right)=\sum_{t=1}^T\left(D_\varphi(\boldsymbol{\theta}_t||\boldsymbol{\theta}_{t-1})-\langle\bold{u}_t,\bold{x}_t\rangle\right).$$

Thus, it follows that 
$$\mathcal{R}(\bold{x})-\langle\boldsymbol{\theta}_T,\bold{x}\rangle\geq -\eta\varphi(\bold{0})-\sum_{t=1}^T\left(D_\varphi(\boldsymbol{\theta}_t||\boldsymbol{\theta}_{t-1})-\langle\bold{u}_t,\bold{x}_t\rangle\right).$$

Combining Lemmas \ref{Positive_R_Bound_D} and \ref{Bregman_Bound}, the previous inequality can be rewritten as
$$\sum_{t=1}^T\langle\bold{x}-\bold{x}_t,\bold{u}_t\rangle\leq \mathcal{R}(\bold{x})+\eta\varphi(\bold{0})+\sum_{t=1}^TD_\varphi(\boldsymbol{\theta}_t||\boldsymbol{\theta}_{t-1})\leq \eta\varphi(\bold{0})+{L\over 2\eta}Tu^2_{max}.$$

Because the previous inequality holds for all $\bold{x}\in \Delta_N$ we conclude:
$$\textsc{R}_{SSA}^T\leq \eta\varphi(\bold{0})+{L\over2\eta}Tu^2_{max}.$$
\eproof
\subsection{Proof of Theorem \ref{SS_Surplus_algorithm_regret}} The bound (\ref{Bound1}) follows from Lemma \ref{Bound_Regret_FTRL}. To derive Eq. (\ref{Bound2}), define the function $\psi(\eta)=\eta\varphi(\bold{0})+\frac{L}{2\eta}Tu^2_{max}$. Given the strict convexity of $\psi(\eta)$, the first order conditions are necessary and sufficient for  a minimum. In particular, we get $$\psi^\prime(\eta)=\varphi(\bold{0})-\frac{L}{2\eta^2}Tu^2_{max}=0$$
The optimal $\eta$ is given by $\eta^*=\sqrt{{LTu^2_{max}\over2\varphi(\bold{0})}}$. Then, it follows that $\psi(\eta^*)=u_{max}\sqrt{2\varphi(\bold{0})LT}$. Thus  we conclude that

$$\textsc{R}_{SSA}^T\leq \psi(\eta^*)=2u_{max}\sqrt{\varphi(\bold{0})LT}.$$\eproof
\subsection{Proof of Theorem \ref{SS_algorithm_GEV}} Combining Lemma \ref{Result_GEV} with Lemma \ref{Bound_Regret_FTRL}
we obtain the bound (\ref{Regret_GEV_SSA}). Following the argument used in proving Theorem \ref{SS_Surplus_algorithm_regret} combined with $L={2M+1\over \eta}$ we obtain the optimized regret bound  (\ref{Regret_GEV_SS}).\eproof

\begin{lemma}\label{GNL_regret_lemma} In the GNL model the following statements hold:
	\begin{itemize}
		\item[i)] The Social Surplus function has a Lipschitz continuous gradient with constant  $({2\over \min_{k}\lambda_k}-1)/\eta.$
		\item[ii)] $\log G(\bold{1})=\log \sum_{k=1}^K\left(\sum_{i=1}^N\alpha^{1/\lambda_k}_{ik} \right)^{\lambda_k}\leq \log N,$
	\end{itemize} 
	
\end{lemma}
\proof i) This follows from a direct application of  \citet[Cor. 4]{Nesterovetal2019}. ii) To prove this, we first show that  for $\lambda_k<\lambda_k^\prime$ we have:

$$\left(\sum_{i=1}^N\alpha^{1/\lambda_k}_{ik} \right)^{\lambda_k}\leq \left(\sum_{i=1}^N\alpha^{1/\lambda^\prime_k}_{ik} \right)^{\lambda^\prime_k}\quad\mbox{for $k=1,\ldots,K.$}$$
Let $p_k={1\over \lambda_k}$ and $p_{k^\prime}={1\over \lambda_{k^\prime}}$, noting that $p^\prime_k<p_k$ whenever $\lambda_k<\lambda_k^\prime$. Using this change of variable, we can write the following ratio

\begin{eqnarray}
{\left(\sum_{i=1}^N\alpha^{p_k}_{ik} \right)^{1/p_k}\over \left(\sum_{j=1}^N\alpha^{p^\prime_k}_{jk} \right)^{1/p^\prime_k}}&=& \left({\sum_{i=1}^N\alpha^{p_k}_{ik} \over \left(\sum_{j=1}^N\alpha^{p^\prime_k}_{jk} \right)^{p_k/p^\prime_k}}\right)^{1/p_k}\nonumber \\
&=& \left(\sum_{i=1}^N\left({\alpha^{p^\prime_k}_{ik} \over \sum_{j=1}^N\alpha^{p^\prime_k}_{jk}}\right)^{p_k/p_k^\prime}\right)^{1/p_k}\nonumber\\
&\leq& \left(\sum_{i=1}^N\left({\alpha^{p^\prime_k}_{ik} \over \sum_{j=1}^N\alpha^{p^\prime_k}_{jk}}\right)\right)^{1/p_k}=1.\nonumber
\end{eqnarray}
The last inequality implies that 
$$\left(\sum_{i=1}^N\alpha^{1/\lambda_k}_{ik} \right)^{\lambda_k}\leq \left(\sum_{i=1}^N\alpha^{1/\lambda^\prime_k}_{ik} \right)^{\lambda^\prime_k}\quad\mbox{for $k=1,\ldots,K.$}$$
Then for $\lambda_k=1$ for $k=1,\ldots, K$, we get
$$\log \sum_{k=1}^K\left(\sum_{i=1}^N\alpha^{1/\lambda_k}_{ik} \right)^{\lambda_k}\leq \log \sum_{k=1}^K\sum_{i=1}^N\alpha_{ik}=\log N.$$
\eproof

\subsection{Proof of Proposition \ref{GNL_regret}} Combining Lemma \ref{GNL_regret_lemma} with Theorem \ref{SS_Surplus_algorithm_regret}, the conclusion follows at once.\eproof

\subsection{Proof of Proposition \ref{R_smooth}}
i) From the definition of $\mathcal{R}(\bold{x})$ we know that $\mathcal{R}(\bold{x})=\eta\varphi^\ast(\bold{x};\eta)$
where $\varphi^\ast(\bold{x};\eta)$ is the convex conjugate of  the parametrized social surplus function $\varphi(\boldsymbol{\theta}/\eta)$. By Lemma \ref{Gradient_Lipschitz}  it follows  that $\varphi(\boldsymbol{\theta}/\eta)$ is $L$-Lipschitz continuous.  Applying  Lemma \ref{Baillon_Haddad} it follows that $\varphi^\ast(\bold{x};\eta)$ is $1/L$ strongly convex. Then it follows that $\mathcal{R}(\bold{x})$ is ${\eta\over L}$-strongly convex. \\
ii) This is a direct implication of Lemma \ref{Convexity_smoothness}.\\
iii) It is easy to see that $ \langle \boldsymbol{\theta},\bold{x}\rangle-\mathcal{R}(\bold{x})$ is a  ${\eta\over L}$-strongly concave function on $\Delta_N$. This implies that  the optimization problem $\max_{\bold{x}\in \Delta_N}\{\langle \boldsymbol{\theta},\bold{x}\rangle-\mathcal{R}(\bold{x})
\}$ must have a unique solution. Let $\bold{x}^*$  be the unique optimal solution. Using the differentiability of $\mathcal{R}(\bold{x})$ combined with the Fenchel equality it follows that $\bold{x}^*=\nabla\varphi(\boldsymbol{\theta})$ iff   $\nabla\varphi(\boldsymbol{\theta})=\arg\max_{\bold{x}\in \Delta_N}\{\langle\boldsymbol{\theta},\bold{x}\rangle-\mathcal{R}(\bold{x})\}$.\eproof
\subsection{Proof of Theorem \ref{Equiv_SSA_FTRL}} From Proposition \ref{R_smooth} we know  $\nabla\varphi(\boldsymbol{\theta}_t)=\bold{x}_{t+1}=\arg\max_{\bold{x}\in \Delta_N}\{\langle\boldsymbol{\theta}_t,\bold{x}\rangle-\R(\bold{x})\}$. Plugging in this observation in the FTRL algorithm the equivalence follows at once.\eproof

\subsection{Proof of Corollary \ref{No_regret_Discrete_choice}} From Proposition \ref{R_smooth}iii) we know $$\bold{x}_{t+1}=\nabla\varphi(\boldsymbol{\theta}_{t})=\arg\max_{\bold{x}\in\Delta_N}\left\{\langle\boldsymbol{\theta}_t,\bold{x}\rangle-\R(\bold{x})\right\}.$$
 
 This fact implies that  the FRTL algorithm is equivalent to the SSA.  Then the argument used in proving Theorem \ref{SS_Surplus_algorithm_regret} applies. Thus $\textsc{R}^T_{FTRL}$ is bounded by the same term that bounds $\textsc{R}^T_{SSA}$. Similarly, the same optimized bound achieved in $\textsc{R}^T_{SSA}$ applies to $\textsc{R}^T_{FTRL}$. \eproof

\subsection{Proof of Proposition \ref{Recursive_Choice_FTRL}} Let us focus at period $t+1$. Accordingly, the associated  Lagrangian is given by:
 	$$\mathcal{L}(\bold{x}_{t+1};\lambda)=\sum_{i=1}^N\boldsymbol{\theta}_{it} \bold{x}_{it+1}-{\eta}\sum_{i=1}^N\bold{x}_{it+1}\log\Phi_i(\bold{x}_{t+1})+\lambda\left(\sum_{i=1}^N\bold{x}_{it+1}-1\right).$$
 	From \cite[Prop. A1ii)]{Fosgerauetal2019a} we know that $\Phi(\bold{x}_{t+1})$ is differentiable with 
 	$$\sum_{j=1}^N\bold{x}_{jt+1}{\partial \Phi_j(\bold{x}_{t+1})\over \partial \bold{x}_{it+1}}=1\quad\forall i\in A.$$
 	Using this fact, the set of  first order conditions can be written as:
 	\begin{eqnarray}
 		{\partial \mathcal{L}(\bold{x}_{t+1};\lambda)\over \partial\bold{x}_{t+1}}&=&\boldsymbol{\theta}_t-{ \eta}\log \Phi(\bold{x}_{t+1})-{ \eta}+\lambda=0.\label{FOC11}\\
 		{\partial \mathcal{L}(\bold{x}_{t+1};\lambda)\over \partial\lambda}&=&\sum_{i=1}^n\bold{x}_{it+1}-1=0.\label{FOC22}
 	\end{eqnarray}
 	Noting that  $\boldsymbol{\theta}_t=\bold{u}_t+\boldsymbol{\theta}_{t-1}$,  Eq. (\ref{FOC11}) can be expressed as
 	\begin{eqnarray}\label{FOC33}
 		e^{\bold{u}_t/\eta+\boldsymbol{\theta}_{t-1}/\eta}e^{\lambda/\eta-1}&=&\Phi(\bold{x}_{t+1})
 	\end{eqnarray}
 	Recalling that $\Phi(\cdot)=H^{-1}(\cdot)$, from (\ref{FOC33}) we get:
 	\begin{eqnarray}
 		H(e^{\bold{u}_t/\eta+\boldsymbol{\theta}_{t-1}/\eta}e^{\lambda/\eta-1})&=&\bold{x}_{t+1}.\nonumber
 	\end{eqnarray}
 	Noting that $H(\cdot)$ is homogeneous of degree 1, we get:
 	\begin{eqnarray}
 		H(e^{\bold{u}_t/\eta+\boldsymbol{\theta}_{t-1}/\eta})e^{\lambda/\eta-1}&=&\bold{x}_{t+1}.\nonumber
 	\end{eqnarray}
 	Using (\ref{FOC22}) we find that 
 	\begin{eqnarray}
 		e^{\lambda/\eta-1}&=&{1\over \sum_{j=1}^NH_j(e^{\bold{u}_t/\eta+\boldsymbol{\theta}_{t-1}/\eta})}.\nonumber
 	\end{eqnarray}
 	Then it is easy to see that 
 	\begin{equation}\label{FOC44}
 		\bold{x}_{jt+1}={H_j(e^{\bold{u}_t/\eta+\boldsymbol{\theta}_{t-1}/\eta})\over \sum_{i=1}^NH_i(e^{\bold{u}_t/\eta+\boldsymbol{\theta}_{t-1}/\eta})}\quad \mbox{for all  $j\in A, t\geq 1.$}
 	\end{equation}
 	From (\ref{FOC44}), it is easy to see that at period $t$ we must have:
 	\begin{equation}\label{FOC55}
 		\bold{x}_{jt}={H_j(e^{\boldsymbol{\theta}_{t-1}/\eta})\over \sum_{i=1}^NH_i(e^{\boldsymbol{\theta}_{t-1}/\eta})}\quad \mbox{for all  $j\in A$.}
 	\end{equation}
 	Once again, using the fact that $\Phi(\cdot)$ is homogeneous of degree 1,  in (\ref{FOC55}) we find:
 	\begin{equation}\label{FOC51}
 		\Phi(\bold{x}_{t})\sum_{i=1}^NH_i(e^{\boldsymbol{\theta}_{t-1}/
 			\eta})=e^{\boldsymbol{\theta}_{t-1}/\eta}\quad \mbox{for  all $j\in A$.}
 	\end{equation}
 	Define $w_{t-1}\triangleq\log\left(\sum_{i=1}^NH_i(e^{\boldsymbol{\theta}_{t-1}/\eta})\right)$. Using this definition, combined with Eq. (\ref{FOC44}),  we get:
 	\begin{equation}
 		\bold{x}_{jt+1}={H_j(\Phi(\bold{x}_{t})e^{\bold{u}_t/\eta+ w_{t-1}})\over\sum_{i=1}^N H_i(\Phi(\bold{x}_{t})e^{\bold{u}_t/\eta+w_{t-1}})}.\nonumber
 	\end{equation}
 	Finally, using the homogeneity of $H$, we conclude that 
 	\begin{equation}
 		\bold{x}_{jt+1}={H_j(e^{\bold{u}_t/\eta+\alpha(\bold{x}_t)})\over\sum_{i=1}^N H_i(e^{\bold{u}_t/\eta+\alpha(\bold{x}_t)})}, \quad\forall j\in A, t\geq 1.\nonumber
 	\end{equation}
 	\eproof 

    \subsection{Proof of Proposition \ref{OFTRL_Regret}} This result follows from combining Lemmas \ref{OFTRL_Regret} and \ref{Cases_regret} and optimizing over $\eta.$


\vspace{2ex}

\vspace{2ex}
\newpage

\section{Online material not for publication}\label{Appendixb}
\begin{lemma}\label{Bound_OFTRL}Let Assumptions \ref{Shocks_Assumption} and \ref{Gradient_LL} hold.  Then in the OFTRL algorithm the following hold:
	\begin{equation}\label{Regret_OFTRL}
	\sum_{t=1}^T\langle\bold{x}^*-\bold{x}_t,\bold{u}_t\rangle\leq \eta\varphi(\bold{0})+{L\over 2\eta}\sum_{t=1}^{T}\left\|\mathbf{u}_{t}-\boldsymbol{\beta}_{t}\right\|_{*}^{2}.
	\end{equation}
\end{lemma}
\proof The proof of this Lemma follows from a simple adaptation of \cite[Lemma 2]{RakhlinSridharan2013}.\eproof

\begin{lemma}\label{Cases_regret}In the OFTRL with $S$-step recency bias, the following inequality holds:
		$$\sum_{t=1}^T\|\bold{u}_t-\boldsymbol{\beta}_t\|^2_\ast\leq S^2\sum_{t=1}^T\|\bold{u}_t-\bold{u}_{t-1}\|^2_\ast$$
		
	
\end{lemma}

\proof In proving this result we follow the proof of Lemma 21 in \citet{NIPS2015_5763}. In particular, we use the following:

\begin{eqnarray}
\sum_{t=1}^T\|\bold{u}_t-\boldsymbol{\beta}_t\|^2_\ast&=& \sum_{t=1}^{T}\left\|\mathbf{u}_{t}-\frac{1}{S} \sum_{\tau=t-H}^{t-1} \mathbf{u}_{\tau}\right\|_{*}^{2},\nonumber\\
&=&  \sum_{t=1}^{T}\left(\frac{1}{S} \sum_{\tau=t-S}^{t-1}\left\|\mathbf{u}_{t}-\mathbf{u}_{\tau}\right\|_{*}\right)^{2}.\label{Eq.S.Steps}
\end{eqnarray}

By the triangle inequality we get
\begin{eqnarray}
\frac{1}{S} \sum_{\tau=t-S}^{t-1}\left\|\mathbf{u}_{t}-\mathbf{u}_{\tau}\right\|_{\ast}&\leq & \frac{1}{S} \sum_{\tau=t-S}^{t-1} \sum_{q=\tau}^{t-1}\left\|\mathbf{u}_{q+1}-\mathbf{u}_{q}\right\|_{*}\nonumber\\
&=& \sum_{\tau=t-S}^{t-1} \frac{t-\tau}{S}\left\|\mathbf{u}_{\tau+1}-\mathbf{u}_{\tau}\right\|_{*} \leq \sum_{\tau=t-S}^{t-1}\left\|\mathbf{u}_{\tau+1}-\mathbf{u}_{\tau}\right\|_{*}\nonumber
\end{eqnarray}

By the Cauchy-Schwarz inequality, we get:
$$
\left(\sum_{\tau=t-S}^{t-1}\left\|\mathbf{u}_{\tau+1}-\mathbf{u}_{\tau}\right\|_{*}\right)^{2} \leq S \sum_{\tau=t-S}^{t-1}\left\|\mathbf{u}_{\tau+1}-\mathbf{u}_{\tau}\right\|_{*}^{2}.
$$
Thus it follows that:
\begin{eqnarray}
\sum_{t=1}^T\|\textbf{u}_t-\boldsymbol{\beta}_t\|^2_\ast&\leq& S\sum_{t=1}^{T} \sum_{\tau=t-S}^{t-1}\left\|\mathbf{u}_{\tau+1}-\mathbf{u}_{\tau}\right\|_{*}^{2}\nonumber\\
&\leq& S^{2} \sum_{t=1}^{T}\left\|\mathbf{u}_{t}-\mathbf{u}_{t-1}\right\|_{*}^{2}\nonumber
\end{eqnarray}

\eproof

\vspace{2ex}

\subsection{Applications of the GNL model}\label{AppendixGNL}

\subsubsection{The Nested Logit (NL) model}\label{Applications_GNL} 
The NL model proposed by \citet{McFadden1978} as a particular case of the GNL in the sense that each alternative $i\in A$ belongs to a unique nest. In other words, the NL is a model in which the nests are 
\emph{mutually exclusive}. In particular,  each allocation parameter  $\alpha_{ik}$ is: 
$$\alpha_{ik}=\begin{cases} 1 &\mbox{if alternative $i\in \mathcal{N}_k$ }  \\ 
	0& \mbox{otherwise}  \end{cases} .
$$

Accordingly,  the  generator $G$ corresponds to:
$$G(e^{\boldsymbol{\theta}_t/\eta})=\sum_{k=1}^K\left(\sum_{i\in \mathcal{N}_k}(e^{\boldsymbol{\theta}_{it}/\eta})^{	1/\lambda_k}\right)^{\lambda_k}.$$

In this model the gradient $\nabla\varphi(\boldsymbol{\theta})$ is  Lipschitz continuous with constant $
\left({2\over\min\lambda_k}-1\right) /\eta$. Thus, by Proposition \ref{GNL_regret} we conclude that the NL model is Hannan consistent.
\smallskip

It is worth mentioning that when $\lambda_k=1$ for $k=1,\ldots,K$, the NL boils down to the MNL.\footnote{We recall that when $\lambda_k=1$ for $k=1,\ldots,N$  the elements of the  random vector $\epsilon_t$ are independent. } Under this parametrization, the generator $G$ can be written as
\begin{eqnarray}
	G(e^{\boldsymbol{\theta}_t/\eta})&=&\sum_{k=1}^K\sum_{i\in \mathcal{N}_k}e^{\boldsymbol{\theta}_{it}/\eta},\nonumber\\
	&=&\sum_{i=1}^Ne^{\boldsymbol{\theta}_{it}/\eta}.\nonumber
\end{eqnarray}
Thus, as a direct result of Proposition \ref{GNL_regret}, we find that the MNL is Hannan consistent.

\subsubsection{The Cross Nested Logit (CNL) model} \citet{Vosha1997} introduces the
cross nested logit model. The main assumption of this model is that $\lambda_k=\lambda$ for all $k=1,\ldots,K$. Thus, the generator $G$ boils down to the expression: $$G(e^{\boldsymbol{\theta}_t/\eta})=\sum_{k=1}^K\left(\sum_{i=1}^N\left(\alpha_{ik}e^{\boldsymbol{\theta}_{it}/\eta}\right)^{1/\lambda}\right)^\lambda.$$
In addition,  in this case the  constant $M$ is given by $M={2\over \lambda}-1$. Accordingly, the Social Surplus function has Lipschitz continuous gradient with constant $({2\over \lambda}-1)/\eta.$
\subsubsection{The Paired Combinatorial Logit (PCL) model} In this model each pair of alternatives is represented by a nest. Formally, the set of nests is defined as $\mathcal{N}=\left\{(i,j)\in A\times A: i\neq j\right \}$.  Accordingly, we  define 
$$\alpha_{ik}=\begin{cases} {1\over 2(N-1)} &\mbox{if $k=(i, j),(j, i)$ with $j \neq i$ }  \\ 
0& \mbox{otherwise}  \end{cases} .
$$

Using the previous expression, the  generator $G$ can be written as:
$$G(e^{\boldsymbol{\theta}_t/\eta})=\sum_{k=(i,j)\in \mathcal{N}}\left((\alpha_{ik}e^{\boldsymbol{\theta}_{it}/\eta})^{1/\lambda_k}+(\alpha_{jk}e^{\boldsymbol{\theta}_{jt}/\eta})^{1/\lambda_k}\right)^{\lambda_k}.$$

In this case   the Lipschitz constant is  $({2\over \min\lambda_k}-1)/\eta.$

 In addition, $\log G(\bold{1})\leq\log N$. Then the regret analysis follows from Proposition \ref{GNL_regret}.
\subsubsection{The Ordered GEV (OGEV) model} \citet{Small1987} studies a GEV model in which the alternatives  are allocated to nests based on their \emph{proximity} in an ordered set.  Following \citet{Small1987} we define the set of overlapping nests to be
$$\mathcal{N}=\{1,\ldots,N+N^\prime\},$$
with $\alpha_{i\ell}>0$ for all $\ell\in\{i,\ldots,N+N^\prime\}$ and $\alpha_{i\ell}=0$ for $i\in\mathcal{N}\setminus \{i,\ldots,N+N^\prime\}$, and each alternative lies exactly in $N^\prime+1$ of these nests. 
In the model there are $N+N^\prime$ overlapping nests. Each nest $\ell\in \mathcal{N}$ is defined as $\mathcal{N}_\ell=\{i\in A:l-m\leq i\leq \ell \}$
where $i\in\mathcal{N}_l$ for $\ell=i,\ldots,i+N^\prime$. Despite this rather complex description, the generator function $G$ takes the familiar form:
$$G(e^{\boldsymbol{\theta}_t/\eta})=\sum_{k=1}^{N+N^\prime}\left(\sum_{i\in \mathcal{N}_k}(\alpha_{ik}e^{\boldsymbol{\theta}_{it}/\eta})^{	1/\lambda_k}\right)^{\lambda_k}$$
Thus in this case  the Lipschitz constant is given by $({2\over \min_{k=1,\ldots,K}\lambda_k}-1)/\eta$. Moreover,  $\log G(\bold{1})\leq\log N$. Thus Proposition \ref{GNL_regret} applies and we conclude that the  OGEV model is Hannan consistent.
\subsubsection{ Principles of  Differentiation GEV model (PDGEV)}\citet{Bresnahanetal1997} introduce  the PDGEV model. This appproach is based on the idea of  markets for differentiated products. Using this idea, the set  of nests is defined in terms of the attributes that characterize the different products (goods). For instance, in  the context of transportation modeling, the attributes can include  mode to work, destination, number of cars, and residential location. Accordingly, let $D$ be the set of attributes with $\mathcal{N}=\bigcup_{d\in D}\mathcal{N}_d$  and $\mathcal{N}_d=\{k\in \mathcal{N}:\mbox{nest $k$  contains products with attribute $d$}\}$ be the nest that contains the alternatives with attribute $d.$ Similarly, let $\mathcal{N}_{kd}$ denote the nest $k$ with attribute $d$.
$$\alpha_{ik}=\left\{\begin{array}{cl}\alpha_{d} & \text { if } i \in N_{kd} \text { and } k \in \mathcal{N}_{d} \\ 0 & \text { otherwise}\end{array}\right.$$
\smallskip

In this case the  generator $G$ takes the form:
$$G(e^{\boldsymbol{\theta}_t/\eta})=\sum_{d\in D}\alpha_d\sum_{k\in \mathcal{N}_d}\left(\sum_{i\in \mathcal{N}_{kd}}e^{\boldsymbol{\theta}_{it}/\eta\lambda_d}   \right)^{\lambda_d}.$$

It is easy to see that the previous generator is a particular case of the GNL model. Furthermore, the Lipschitz constant is  $({2\over min_{d=1,\ldots,D}\lambda_d}-1)/\eta$ and $\log G(\bold{1})\leq\log N$. Thus  Proposition \ref{GNL_regret} applies in a direct way.
\smallskip

We close this appendix summarizing the regret bounds  for  the models  described  in the main text and  in this appendix. Table \ref{Summary} below makes explicit our regret analysis. The table displays how several GEV models shares  the same optimized regret bound.
\begin{table}[h!]
	\begin{center}
		
		\begin{tabular}{l|c|r} 
			\textbf{Model} & \textbf{Optimal $\eta$} & \textbf{Regret Bound}\\
			\hline
			RUM & $\sqrt{{LTu_{max}^2\over 2\varphi(0)}}$ & $u_{max}\sqrt{2\varphi(0)LT}$\\
			GEV & $\sqrt{{(2M+1)Tu_{max}^2\over 2\log G(1)}}$ & $u_{max}\sqrt{2 \log G(\textbf{1})(2M+1)T}$\\
			GNL & $\sqrt{\left( {2\over \min_{k}\lambda_k}-1\right)Tu^2_{max}\over 2\log N}$ & $u_{max}\sqrt{2\log N \left( {2\over \min_{k}\lambda_k}-1\right)T}$\\
			PCL&$\sqrt{\left( {2\over \min_{k}\lambda_k}-1\right)Tu^2_{max}\over 2\log N}$  & $u_{max}\sqrt{2\log N \left( {2\over \min_{k}\lambda_k}-1\right)T}$\\
			CNL&$\sqrt{\left( {2\over \lambda}-1\right)Tu^2_{max}\over 2\log N}$ &  $u_{max}\sqrt{2\log N \left( {2\over \lambda}-1\right)T}$\\
			
			OGEV&$\sqrt{\left( {2\over \min_{k}\lambda_k}-1\right)Tu^2_{max}\over 2\log N}$ & $u_{max}\sqrt{2\log N \left( {2\over \min_{k}\lambda_k}-1\right)T}$\\
			PDGEV&$\sqrt{\left( {2\over \min_{d}\lambda_d}-1\right)Tu^2_{max}\over 2\log N}$ & $u_{max}\sqrt{2\log N \left( {2\over \min_{d}\lambda_d}-1\right)T}$\\
			NL& $\sqrt{\left( {2\over \min_{k}\lambda_k}-1\right)Tu^2_{max}\over 2\log N}$ & $u_{max}\sqrt{2\log N \left( {2\over \min_{k}\lambda_k}-1\right)T}$\\
			Logit&$\sqrt{Tu^2_{max}\over 2\log N}$ & $u_{max}\sqrt{2\log N T}$

		\end{tabular}
		\vspace{4ex}
		\caption{Summary of the optimized regret bound for the RUM and several GEV models.}
		\label{Summary}
		
	\end{center}
\end{table}


\end{document}